\documentclass[12pt]{iopart}
\usepackage[symbol]{footmisc}

\expandafter\let\csname equation*\endcsname\relax
\expandafter\let\csname endequation*\endcsname\relax
\usepackage{amsmath,mathtools}
\usepackage{amsfonts}
\usepackage{amssymb}
\usepackage{physics}
\usepackage{fancyhdr}
\usepackage{tikz}
\usepackage{tikz-feynman}
\usetikzlibrary{decorations.markings}
\usepackage{parskip}
\usepackage{lmodern}
\usepackage{hyperref}
\usepackage[normalem]{ulem}
\usepackage{comment}
\usepackage{bbold}
\newcommand{\citep}[1]{\cite{#1}}
\newcommand{\plaind}{\mathrm{d}}
\newcommand{\dint}[1]{\mathchoice{\!\plaind#1\,}{\!\plaind#1\,}{\!\plaind#1\,}{\!\plaind#1\,}}

\newcommand{\dTWOint}[1]{\ddintx{#1}{2}}

\newcommand{\ddintx}[2]{\mathchoice{\!\plaind^{#2}#1\,}{\!\plaind^{#2}#1\,}{\!\plaind^{#2}#1\,}{\!\plaind^{#2}#1\,}}

\usepackage[T1]{fontenc}
\newcommand{\dbar}{\text{\dj}}
\newcommand{\deltabar}{\delta\mkern-6mu\mathchar'26}
\newcommand{\dintbar}[1]{\mathchoice{\!\dbar#1\,}{\!\dbar#1\,}{\!\dbar#1\,}{\!\dbar#1\,}}

\newcommand{\dTWOintbar}[1]{\mathchoice{\!\dbar^2#1\,}{\!\dbar^2#1\,}{\!\dbar^2#1\,}{\!\dbar^2#1\,}}

\newcommand{\Dint}[1]{\mathcal{D}\!#1\,}

\usepackage{dsfont}

\newcommand{\gpset}[1]{\mathds{#1}}

\newcommand{\canetset}[1]{{\mathchoice {\hbox{$\sf\textstyle #1\kern-0.4em #1$}}
{\hbox{$\sf\textstyle #1\kern-0.4em #1$}}
{\hbox{$\sf\scriptstyle #1\kern-0.3em #1$}}
{\hbox{$\sf\scriptscriptstyle #1\kern-0.2em #1$}}}}

\newcommand{\Rset}{\gpset{R}}

\newcommand{\gpvec}[1]{\mathbf{#1}}

\newcommand{\zerovec}{\gpvec{0}}
\newcommand{\nullvec}{\zerovec}

\newcommand{\evec}{\gpvec{e}}

\newcommand{\kvec}{\gpvec{k}}

\newcommand{\xvec}{\gpvec{x}}

\newcommand{\etavec}{\boldsymbol{\eta}}

\newcommand{\transpose}{\mathsf{T}}

\newcommand{\ave}[2][]{\mathchoice%
{\left\langle #2 \right\rangle_{#1}}%
{\langle #2\rangle_{#1}}
{\langle #2\rangle_{#1}}%
{\langle #2\rangle_{#1}}}

\newcommand{\EXP}[1]{\operatorname{exp}\!\Bigg\{#1\Bigg\}}

\renewcommand{\exp}[1]{\mathchoice%
{\mathrm{e}^{#1}}%
{\operatorname{exp}(#1)}
{\operatorname{exp}\left(#1\right)}%
{\operatorname{exp}\left(#1\right)}}

\usepackage{xspace}
\newcommand{\latin}[1]{{\it #1}}

\newcommand{\elabel}[1]{\label{#1}}
\newcommand{\Erefs}[1]{Equations~\eref{#1}}

\newcommand{\prob}{\mathcal{P}}
\newcommand{\chitilde}{\tilde{\chi}}
\newcommand{\chidagger}{\chi^\dagger}
\newcommand{\action}{\mathcal{A}}
\newcommand{\actionPert}{\action_{\text{\tiny pert}}}
\newcommand{\actionHarm}{\action_0}
\newcommand{\aveHarm}[1]{\ave{#1}_0}
\newcommand{\imag}{\mathring{\imath}}

\usepackage{mathtools}
\newcommand\underrel[3][]{\mathrel{\mathop{#3}\limits_{%
			\ifx c#1\relax\mathclap{#2}\else#2\fi}}}

\begin{document}

\title[Field theory of transiently chiral active particles]{Looking Back: Field theory of transiently chiral active particles}

\author{Callum Britton$^1$, Gunnar Pruessner$^1$ and Thibault Bertrand$^1$}
\address{$^1$\it{Department of Mathematics, Imperial College London, 180 Queen’s Gate, London SW7 2AZ}}
\ead{callum.britton19@imperial.ac.uk, g.pruessner@imperial.ac.uk, t.bertrand@imperial.ac.uk}

\begin{abstract}
We derive a Doi-Peliti Field Theory for transiently chiral active particles in two dimensions, that is, active Brownian particles that undergo tumbles via a diffusing reorientation angle. Using this framework, we compute the mean squared displacement for both uniformly distributed and fixed initial reorientations. We also calculate an array of orientation-based observables, to quantify the transiently chiral behaviour observed.
\end{abstract}

\section{Introduction}
\label{sec:intro}

Life is made of fundamentally out-of-equilibrium processes, where maintaining homeostasis and performing biological functions requires a constant flux of energy. At equilibrium, a state synonymous with thermodynamic death, living systems cease to operate. The study of active matter \cite{Cates:2012,Bechinger2016} provides a robust framework for examining these far-from-equilibrium systems, where individual units consume and dissipate energy locally. Active matter thus represents a new paradigm of non-equilibrium physics, offering quantitative tools to describe complex biological processes, including systems such as cells crawling on a substrate \cite{SchnyderETAL:2017,BotheETAL:2025} or flagellated bacteria like \latin{Escherichia coli} (\latin{E.~coli}) \cite{KurzthalerETAL:2024}. In active systems, the local energy consumption is translated into mechanical forces exerted on their surroundings; most often, this mechanical work translates to self-propulsion.

The last decades have seen the development and analysis of minimal models aiming to capture some of the biological complexity \cite{Cates:2012}. Two of the most established models are the Active Brownian Particles (ABPs) \cite{HowseETAL:2007} and run-and-tumble particles (RnTs) \cite{TailleurCates:2008,CatesTailleur:2013}, both of which have provided critical insights into phenomena such as bacterial motility and transport in complex environments. For instance, the motion of \latin{E.~coli} is a direct consequence of the transfer of momentum between its rotating chiral flagella and the surrounding fluid, a process that has been extensively modeled through RnTs \cite{berg-nature-1972,berg-book-1993,berg-book-2004,PattesonETAL:2015}. 

Theoretical approaches to understanding the dynamics of RnTs and ABPs have employed a variety of analytical and computational methods. Direct calculations, such as solving Fokker-Planck equations, provide a detailed description of the stochastic dynamics of these systems, capturing key statistical properties such as mean squared displacement and velocity correlations. Beyond these direct approaches, field-theoretic methods have played a central role in exploring microscopic-level descriptions as well as large-scale behavior and fluctuations. In particular, Doi-Peliti (DP) \cite{Doi:1976, Peliti:1985} and Martin-Siggia-Rose (MSR) formalisms \cite{MartinSiggiaRose:1973,Janssen:1976,DeDominicis:1976} are being used \cite{ZhangETAL:2024,Pruessner:2025} to systematically derive analytic equations of motion of active matter, enabling the study of emergent phenomena such as phase separation, clustering, and collective motion in active systems. These field-theoretic frameworks offer a powerful approach to understanding the non-equilibrium nature of RnTs and ABPs, including rare events \cite{SantraETAL:2020}, fluctuations \cite{BanerjeeETAL:2020,Redig:2024} and entropy production \cite{Pruessner:2025}

Memory effects play a crucial role in bacterial transport, allowing cells to adapt their motion based on past interactions and environmental cues \cite{Friedrich:2007}. A key form of this memory is directional persistence, where bacteria maintain their direction of motion over a period of time before reorienting \cite{Friedrich:2007}. Tuning this persistence is especially important in navigating complex environments, as it allows bacteria to avoid getting trapped and enhances their ability to explore space efficiently \cite{bertrand-prl-2018,Rashid:2019}. Another memory effect lies in the reorientation dynamics itself. In contrast to classical memory-less models like run-and-tumble particles, where successive reorientation events are uncorrelated, experimental evidence shows that correlations between successive reorientation events (or tumbles) may exist \cite{Zhao:2024}. Additionally, chemotaxis relies on the ability of bacteria to remember previous chemical concentrations \cite{BergPurcell:1977,KellerSegal:1971,EndresWingreen:2008}, enabling them to navigate efficiently toward favorable environments \cite{KnightETAL:2024}. Biochemical trail-following behavior in bacteria like \latin{Pseudomonas aeruginosa} (\latin{P.~aeruginosa}) further introduces environmental memory, where cells respond to their own secretions, leading to collective phenomena such as swarming and quorum sensing \cite{MirandaETAL:2022}. In eukaryotes, it was recently shown that motile epithelial cells can modify their environment in ways that leave persistent footprints, effectively giving them a spatial memory that influences their future movement and leads to self-interacting random walk behaviour with broad consequences for their large-scale dynamics \cite{dalessandro-natcomms-2021}.

Incorporating memory into bacterial transport models is essential for capturing the full complexity of biological motility. Directional persistence, observed in confined or crowded environments, introduces short-term memory that enhances space exploration. Collective memory, such as quorum sensing, allows bacterial populations to synchronize behavior based on shared information about population density. Recent theoretical models have begun integrating these memory effects, for example, in the context of models of active particles with a colored noise which introduces correlations between successive reorientation events \cite{Marconi:2015,Potdar:2010}. These models are a first step towards a more accurate framework for understanding bacterial dynamics in complex environments, highlighting the role of memory in optimizing transport and survival. Despite these limited extensions to the now classical ABP and RnT models, our understanding of the importance of memory effects and correlations in the reorientation dynamics of biological systems like \latin{E.~coli} is still crucially lacking. 

In response to this limitation, we introduce a novel model: transiently chiral active particles (TCAPs). These particles, like ABPs, exhibit self-propulsion with diffusing direction of motion but also experience reorientation events with a finite memory, leading to correlations between successive tumbles. Using Doi-Peliti field theory, we derive a number of observables for free TCAPs and establish a flexible framework with which we can easily study extensions of the model.

In \Sref{sec:TCAP}, we introduce the TCAP model and discuss the emergent transient chirality. In \Sref{sec:DPFT}, we derive a field theory for this model. In \Sref{sec:msd}, we calculate and analyse the mean squared displacement starting with the case of a uniform initial $\psi_0$ in \Sref{sec:msdu}, followed by the case of fixed initial $\psi_0$ in \Sref{sec:msdf}. We further calculate a number of orientation-dependent observables including: reorientation angle probability density function in \Sref{sec:marginal}, orientation autocorrelator in \Sref{sec:auto} and position-orientation cross-correlators in \Sref{sec:cross}. \Sref{sec:conclusion} summarises our findings and provides conclusions.

\section{Transiently chiral active particles}
\label{sec:TCAP}

Here, we introduce a novel model of active particles capturing the interplay between active propulsion and correlated stochastic reorientations; we coin this model the transiently chiral active particle (TCAP) model. TCAPs exhibit self-propulsion with a constant speed $v$ in a direction given by an internal degree of freedom $\phi$ and are subject to thermal noise, manifesting through translational diffusion with diffusivity $D_{x}$ and rotational diffusion of its director with diffusivity $D_{\phi}$. In two spatial dimensions, the particle's director is described by a single scalar angle $\phi\in[0,2\pi)$, which evolves stochastically as it undergoes rotational diffusion. 

In addition to the persistent motion by self-propulsion and diffusion, the dynamics are punctuated by discrete, stochastic events termed ``tumbles'' akin to those present in the now classical run-and-tumble particle models \cite{TailleurCates:2008,CatesTailleur:2013}. These tumble events occur randomly and independently with a Poissonian rate $\gamma$, at which point the particle instantaneously reorients itself by an angle $\psi$, i.e. $\phi$ increases by $\psi$. While in the original RnT model, $\psi$ is randomly sampled from a uniform distribution $[0,2\pi)$, so that the directors before and after the tumble are uncorrelated, here we introduce correlations between the reorientations and impose that the increment $\psi\in[0,2\pi)$ is given by another internal degree of freedom that undergoes its own diffusion process, characterized by the diffusion constant $D_{\psi}$. The combination of these mechanisms gives rise to complex dynamics, as the particle explores its environment through a combination of persistent motion, random diffusion, and stochastic reorientation events. 

\subsection{Mathematical description}
\label{sec:model}

In two dimensions, between tumbles, the position of a free TCAP, $\mathbf{r}(t) = (x(t), y(t))^{\transpose}$, evolves according to the following overdamped Langevin equations
\begin{subequations}\elabel{Langevin_eqns}
\begin{align}
	\dot{\mathbf{r}}(t) &= v \, \mathbf{e}(\phi(t)) + \sqrt{2 D_{x}} \, \boldsymbol{\eta}(t) \\
	\dot{\phi}(t) &= \sqrt{2 D_\phi} \, \xi_\phi(t)
\end{align}
\end{subequations}
where $v$ is the constant self-propulsion speed, $\mathbf{e}(\phi(t))$ is the director defined by the orientation angle $\phi(t)$
\begin{align}\label{def_e}
    \mathbf{e}(\phi(t)) = {\cos{(\phi(t))}\choose\sin{(\phi(t))}},
\end{align}
$D_{x}$ is the translational diffusion coefficient and $D_\phi$ is the rotational diffusion coefficient. Furthermore, $\boldsymbol{\eta}\in\Rset^2$ and $\xi_\phi\in\Rset$ are uncorrelated zero-mean, unit-variance Gaussian white noises such that 
\begin{subequations}\elabel{noise_correlators}
\begin{align}
& \ave{\etavec(t)} = \nullvec, \quad \ave{\etavec(t)\etavec^\transpose(t')} = \mathbb{1}_2 \delta(t-t')\\
&\langle \xi_\phi(t) \rangle = 0, \quad \langle \xi_\phi(t) \xi_\phi(t') \rangle =  \delta(t-t')
\end{align}
\end{subequations}
with $\ave{\etavec(t)\etavec^\transpose(t')}$ using the outer product resulting in a $2\times 2$ correlator matrix and $\mathbb{1}_2$ being the $2\times2$ identity matrix.

In the present overdamped regime, where friction dominates, inertial effects are neglected, typical for colloidal systems and biological microswimmers \cite{Purcell:1977}. In the absence of tumbles, the combination of active propulsion along the direction $\mathbf{e}(\phi(t))$ and rotational diffusion by $D_\phi$ leads to persistent motion over a timescale $D_{\phi}^{-1}$ and diffusive behavior beyond that, characterised by an effective diffusivity $D_{\textrm{eff}} = D_{x} + v^2/(2D_\phi)$ \cite{Bechinger2016}.

\begin{figure}[h!]
\centering
\includegraphics[width=\textwidth]{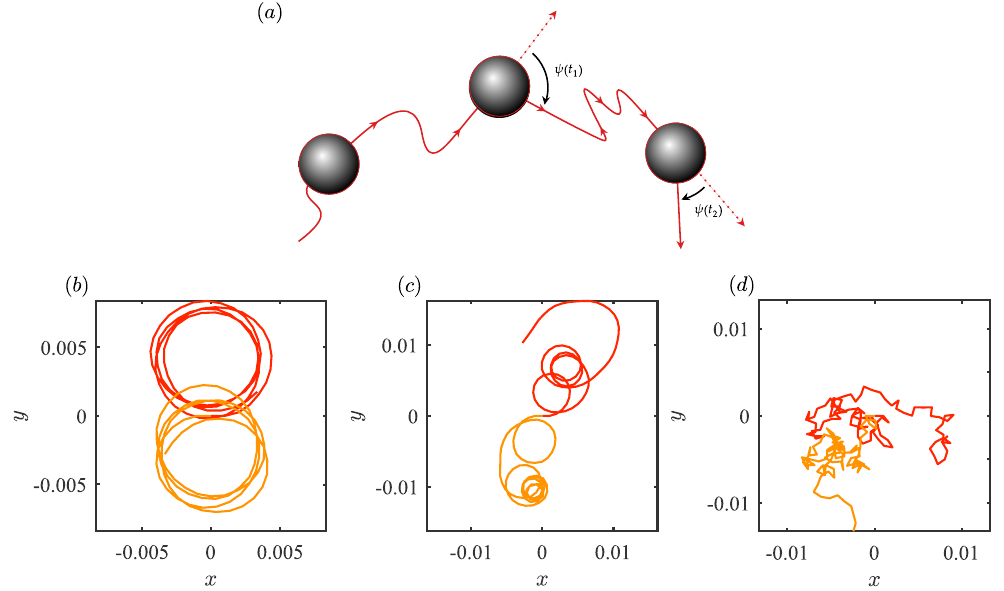}
\caption{Example trajectories for free TCAP particles --- (a) schematic of successive tumbling events with reorientation angles $\{\psi(t_1), \psi(t_2), \ldots\}$. Representative trajectories are shown for two free TCAP with $D_x=0$, $D_\phi=1$, $\gamma=1000$, $v=1$, $\psi_0=\pi/12$ and (b) $D_\psi = 0.0001$, (c) $D_\psi = 1$ and (d) $D_\psi = 10000$.} 
\label{fig:extrajectories}
\end{figure}

In addition to this, {\it tumbles} occur with Poissonian rate $\gamma$, such that the time between tumbles is exponentially distributed with mean $\gamma^{-1}$. When a tumble occurs, the orientation $\phi(t)$ of the TCAP is instantaneously updated to $\phi(t)+\psi(t)$, \Fref{fig:extrajectories}.

The reorientation angle $\psi(t)$ is itself governed by the Brownian dynamics
\begin{equation}
\dot{\psi}(t) = \sqrt{2 D_\psi} \, \xi_\psi(t)
\end{equation}
with $D_\psi$ the diffusivity of the reorientation angle which controls how fast the particle forgets the reorientation angle from any previous tumble event and $\xi_{\psi}(t)$ a zero-mean, unit variance Gaussian white noise,
\begin{equation}
    \ave{\xi_\psi(t)}=0 \quad
    \ave{\xi_\psi(t)\xi_\psi(t')}=\delta(t-t')\ .
\end{equation}
What makes the present system \emph{transiently chiral}, is the persistence of the reorientation angle, so that consecutive tumbles involve correlated, and possibly very similar, reorientation angles.

The motion of a TCAP is thus governed by three timescales: (1) The tumbling timescale $\tau_t=\gamma^{-1}$, the average time between tumbles; (2) The persistence time $\tau_p = 1/(D_\phi + \gamma)$ characterizing the persistence of the self-propelled motion of the TCAP, as it ``forgets'' its orientation, $D_\phi$, or until a tumble occurs, $\gamma$; (3) The correlation time $\tau_{\psi} = 1/D_{\psi}$ of the reorientation angles in successive tumbling events. By tuning these timescales, TCAPs can display qualitatively very different trajectories as shown in \Fref{fig:extrajectories}.

\subsection{Transient chirality}
\label{sec:transient}

In principle, there are six regimes to consider, depending on the relative ordering of the three timescales involved, $\tau_\psi$, $\tau_p$ and $\tau_t$. We exclude from this discussion the translational diffusion by $D_{x}$ which may be thought of as being superimposed. The timescale $\tau_\psi$ of the evolution of the reorientation angle enters only when the tumbles are not drowned out by the angular diffusion $D_\phi$. When the latter is large, tumbles, correlated or not, are barely distinguishable from angular diffusion and TCAPs approach plain ABPs. We therefore assume in the following $D_\phi\ll\gamma$, so that $\tau_t$ and $\tau_p$ represent essentially the same scales. In our reasoning, we will fix the values of $D_{\phi}$ and $\gamma$ 
and thus both the tumbling timescale $\tau_t=1/\gamma$ and the persistence timescale $\tau_p=(D_\phi+\gamma)^{-1}$, 
and consider changing the reorientation timescale $\tau_\psi=1/D_\psi$.

For large enough $\tau_\psi > \tau_t$, the reorientation angle $\psi(t)$ does not have time to significantly change between tumbles, so that successive tumbles occur with roughly the same reorientation angle. As the angular diffusion is low compared to the tumble rate, $D_\phi\ll\gamma$, the motion will consist of a sequence of more or less straight runs with tumbles by about the same angle. This results in \emph{highly correlated chiral motion}, following strikingly ``swirly paths'', \Fref{fig:extrajectories}(b). 
When the reorientation angle $\psi$ is roughly constant and small enough so that a full turn of the director $\evec(\phi(t))$ takes long enough for many tumbles to occur, in the limit $\tau_\psi\gg\tau_t$ the particle will trace a circle of radius $v/(\psi\gamma)$.

Conversely, for small enough $\tau_\psi\ll\tau_t$, the reorientation angle becomes uncorrelated within the typical time between tumbles 
leading to uncorrelated successive reorientation events, \Fref{fig:extrajectories}(d). In this case, the evolution of the reorientation angle has little bearing on the appearance of the trajectories and the behaviour of the TCAP in the limit $\tau_\psi\ll\tau_t$ is that of an RnT combined with an ABP, recovering RnT behaviour as $D_\phi$ vanishes and $D_\psi$ diverges.

For intermediate regimes where $\tau_\psi \approx \tau_t \approx \tau_p$, we expect \emph{transiently chiral motion}, as described in \Sref{sec:marginal} and \Sref{sec:cross}. In the following, we will characterise TCAPs for the full range of parameters.

\section{Doi-Peliti field theory approach}
\label{sec:DPFT}

To cast the Langevin \Erefs{Langevin_eqns} and \eref{noise_correlators} into a field theory, we first rewrite them in Fokker-Planck form,
\begin{multline}\elabel{FPE}
    \partial_t \prob(\xvec,\phi,\psi,t) =
    \big(
    -v \evec(\phi(t)) \cdot \nabla_\xvec + D_{x} \nabla_\xvec^2 + D_\phi \partial_\phi^2 + D_\psi \partial_\psi^2 - r -\gamma
    \big) \prob(\xvec,\phi,\psi,t) \\
    + \gamma \prob(\xvec,\phi-\psi,\psi,t)
\end{multline}
where on the right are, in order of appearance: The advection term stemming from the self-propulsion parameterised by $v$, with director $\evec$, the spatial diffusion, $D_{x}$, the orientational diffusion, $D_\phi$, the diffusion of the reorientation angle, parameterised by $D_\psi$, a mass $r$ to regularise the propagator and make it causal even when $\gamma$ vanishes \cite{Pausch2019}, and finally the tumbling with rate $\gamma$, which has the form of a transmutation, whereby particles ``disappear'' from state $\phi$, which, in turn, is ``fed'' by state $\phi-\psi$ as the angle increases by $\psi$ at a tumble. Even when $\phi$ enters only in periodic functions via $\evec(\phi)$, both $\phi$ and $\psi$ are taken from a periodic domain $[0,2\pi)$, below implemented by a suitable Fourier series.

\Eref{FPE}  immediately translates \cite{Pruessner:2025} to an action for the annihilator field $\chi(\xvec,\phi,\psi,t)$ and its Doi-shifted creator counterpart $\chidagger(\xvec,\phi,\psi,t)=1+\chitilde(\xvec,\phi,\psi,t)$, 
\begin{multline}\elabel{orig_action}
    \action[\chi,\chitilde] = 
    \int_{\Rset^2} \dTWOint{x} \int_{-\infty}^\infty \dint{t}
\int_0^{2\pi} \dint{\phi}
\int_0^{2\pi} \dint{\psi}
\Big\{
\chitilde(\xvec,\phi,\psi,t)
    \big(
    -\partial_t
    -v \evec(\phi(t)) \cdot \nabla_\xvec \\
    + D_{x} \nabla_\xvec^2 + D_\phi \partial_\phi^2 + D_\psi \partial_\psi^2 - r -\gamma
    \big) \chi(\xvec,\phi,\psi,t)
    + 
\gamma 
\chitilde(\xvec,\phi,\psi,t)
\chi(\xvec,\phi-\psi,\psi,t)
\Big\}
\end{multline}
where the spatial integral is over the entire two-dimensional plane $\Rset^2$, similarly the integral over time over the entire real line, whereas $\phi$ and $\psi$ are confined to the interval $[0,2\pi)$.

Using \Eref{orig_action}, observables are calculated via the path integral \cite{Doi:1976,Peliti:1985,TaeuberHowardVollmayr-Lee:2005,Cardy:2008,Pruessner2011,Tauber:2014,Pausch2019}
\begin{equation}\elabel{def_ave}
    \ave{\bullet} = 
    \int \Dint{\chi}\Dint{\chitilde}
    \bullet
    \EXP{\action[\chi,\chitilde]}
\end{equation}
in particular 
\begin{equation}\label{probability_propagator}
\ave{\chi(\xvec,\phi,\psi,t)\chidagger(\xvec_0,\phi_0,\psi_0,t_0)} = 
    \prob(\xvec,\phi,\psi,t; \xvec_0,\phi_0,\psi_0,t_0)\ ,
\end{equation}
the probability density to find at time $t$
a particle at position $\xvec$, with director $\phi$ and reorientation angle $\psi$, having placed exactly one particle (in the vacuum) at time $t_0$ at position $\xvec_0$ with director $\phi_0$ and reorientation angle $\psi_0$.

Within the Doi-Peliti framework \cite{Doi:1976,Peliti:1985}, the action is a complete, exact representation of the Fokker-Planck equation \cite{Pruessner:2025}. The construction of any observable has to respect the commutator relation and subsequently the normal ordering \cite{Cardy:2008,Pruessner2011}, which results in $\chi(\xvec,\phi,\psi,t)$ \emph{generally not} representating a probability density \cite{Cardy:2008} and the exponential in \Eref{orig_action} \textit{not} being a path probability density of the stochastic process of \Erefs{Langevin_eqns} and \eref{noise_correlators}.

The full action as in \Eref{orig_action} 
\begin{equation}
        \action[\chi,\chitilde] = 
    \actionHarm[\chi,\chitilde]  
+
    \actionPert[\chi,\chitilde] 
\end{equation}
firstly consists of the 
bilinear, harmonic part
\begin{equation}\elabel{def_actionHarm}
 \actionHarm[\chi,\chitilde] = 
    \int \dTWOint{x}\, \dint{t}\,
\dint{\phi}\,
\dint{\psi}
\Big\{
\chitilde(\xvec,\phi,\psi,t)
    \big(
    -\partial_t
+ D_{x} \nabla_\xvec^2 + D_\phi \partial_\phi^2 + D_\psi \partial_\psi^2 - r -\gamma
    \big) \chi(\xvec,\phi,\psi,t)
\Big\}
\end{equation}
where we have dropped the integration limits to improve readability.
The path integral over the action \Eref{def_actionHarm} can be taken in closed form, because the operators featuring in the action are easily diagonalised.

The perturbative part
\begin{equation}\elabel{def_actionPert}
    \actionPert[\chi,\chitilde]
    = 
    \int \dTWOint{x}\, \dint{t}\,
\dint{\phi}\,
\dint{\psi}
\Big\{
\chitilde(\xvec,\phi,\psi,t)
\big(
- 
v \evec(\phi(t)) \cdot \nabla_\xvec \chi(\xvec,\phi,\psi,t)
+
\gamma\chi(\xvec,\phi-\psi,\psi,t)
\big)
\Big\}
\end{equation}
absorbs the two terms that are more difficult to handle, namely (some of) the tumbling with rate $\gamma$ that couples degrees of freedom $\phi$ and $\psi$
and the 
self-propulsion which couples degrees of freedom $\xvec$ and $\phi$. Using Mathieu-functions, the latter can be diagonalised as well \cite{ZhangPruessner:2022}, drawing, however, on a  generally less well-characterised eigensystem. As a matter of convenience, we rather use a Fourier eigensystem throughout \cite{ZhangETAL:2024},
\begin{subequations}
\elabel{def_Fourier}
\begin{align}
    \chi(\xvec,\phi,\psi,t) 
    &=\int_{\Rset^2} 
    \dTWOintbar{k}
    \int_{-\infty}^\infty 
    \dintbar{\omega}
    \sum_{n=-\infty}^\infty
    \sum_{m=-\infty}^\infty
    \exp{\imag \kvec\cdot\xvec}
    \exp{-\imag \omega t}
    \exp{\imag n \phi}
    \exp{\imag m \psi}
    \chi_{n,m}(\kvec,\omega)\\
    \chitilde(\xvec,\phi,\psi,t) 
    &=\int_{\Rset^2} \dTWOintbar{k}\int_{-\infty}^\infty \dintbar{\omega}
    \sum_{n=-\infty}^\infty
    \sum_{m=-\infty}^\infty
    \exp{\imag \kvec\cdot\xvec}
    \exp{-\imag \omega t}
    \frac{e^{-\imag n \phi}}{2\pi}
    \frac{e^{-\imag m \psi}}{2\pi}
    \chitilde_{n,m}(\kvec,\omega)    
\end{align}    
\end{subequations}
where momentum $\kvec$ replaces the position $\xvec$, frequency $\omega$ replaces direct time $t$, index $n$ the orientational degree of freedom $\phi$ and index $m$ the reorientation angle $\psi$. The definitions above are a compromise of convenience \cite{Zhang2022b}, as $\chitilde$ draws on the inverse transform for $\phi$ and $\psi$, and tradition, where both $\chi$ and $\chitilde$ use the forward transformation. In the following, barred symbols conveniently absorb powers of $2\pi$, such as $\dbar\omega = d\omega/2\pi$ and $\dbar k = dk/2\pi$, or $\deltabar(\omega) = 2\pi\delta(\omega)$ and $\deltabar(\mathbf{k}) = (2\pi)^{2} \delta(\mathbf{k})$.

For completeness, we state the inverse transforms,
\begin{subequations}
\elabel{def_inverseFourier}
\begin{align}
    \chi_{n,m}(\kvec,\omega)     
    &=
    \int_{\Rset^2} \dTWOint{x} \int_{-\infty}^\infty \dint{t}
    \int_0^{2\pi} \dint{\phi}
    \int_0^{2\pi} \dint{\psi}
    \exp{-\imag \kvec\cdot\xvec}
    \exp{\imag \omega t}
    \frac{e^{-\imag n \phi}}{2\pi}
    \frac{e^{-\imag m \psi}}{2\pi}
    \chi(\xvec,\phi,\psi,t)\\    
    \chitilde_{n,m}(\kvec,\omega) 
    &=
    \int_{\Rset^2} \dTWOint{x} \int_{-\infty}^\infty \dint{t}
    \int_0^{2\pi} \dint{\phi}
    \int_0^{2\pi} \dint{\psi}
    \exp{-\imag \kvec\cdot\xvec}
    \exp{\imag \omega t}
    \exp{\imag n \phi}
    \exp{\imag m \psi}
    \chitilde(\xvec,\phi,\psi,t)\ .
\end{align}    
\end{subequations}

After the transformation, the two parts of the action read
\begin{subequations}
\begin{align}
\elabel{actionHarm_Fourier}
    \actionHarm[\chi,\chitilde]  
    &=-\int_{\Rset^2} \dTWOintbar{k}\int_{-\infty}^\infty \dintbar{\omega}
    \sum_{n=-\infty}^\infty
    \sum_{m=-\infty}^\infty
\chitilde_{n,m}(-\kvec,-\omega)
\\
\nonumber
&\qquad\times
\Big(
-\imag \omega + D_{x} k^2 + D_\phi n^2 + D_\psi m^2 + \gamma + r
\Big)
\chi_{n,m}(\kvec,\omega) \ ,
\\
\elabel{actionPert_Fourier}
    \actionPert[\chi,\chitilde] 
    &=\int_{\Rset^2} \dTWOintbar{k}\int_{-\infty}^\infty \dintbar{\omega}
    \sum_{n=-\infty}^\infty
    \sum_{m=-\infty}^\infty
\chitilde_{n,m}(-\kvec,-\omega)
\\
\nonumber
&\qquad\times\Big(
\frac{v}{2\imag}(k_x + \imag k_y)\chi_{n+1, m}(\mathbf{k},\omega) + \frac{v}{2\imag}(k_x - \imag k_y)\chi_{n-1, m}(\mathbf{k},\omega)
\\
\nonumber
&\qquad\qquad
+
\gamma \chi_{n,m+n}(\kvec,\omega)
\Big)
\end{align}
\end{subequations}
where $\kvec=(k_x,k_y)^\transpose$.
From the harmonic part of the action \Eref{actionHarm_Fourier}, we can read off the bare propagator
\begin{subequations}
    \label{bareprop}
\begin{align} 
     \langle\chi_{n,m}(\mathbf{k},\omega) \tilde{\chi}_{n',m'}(\mathbf{k}',\omega')\rangle_{0} =& \frac{\deltabar(\mathbf{k}+\mathbf{k}')\deltabar(\omega+\omega')\delta_{n,n'}\delta_{m,m'}}{-\imag\omega + D_{x}k^{2} + D_{\phi}n^{2} + D_{\psi}m^{2} + \gamma + r}\\\nonumber
     \equiv& G(k,n,m,\omega)\deltabar(\mathbf{k}+\mathbf{k}')\deltabar(\omega+\omega')\delta_{n,n'}\delta_{m,m'}\\
     \elabel{bare_propagator_feynman}
     \hat{=}& \begin{tikzpicture}[baseline={([yshift=-0.7ex]current bounding box.center)}]
    \begin{feynman}
    \vertex (w) at (0,0);
    \vertex (w0) at (1.5,0);
    \diagram* {(w) -- [red, line width=0.25mm, -] (w0),
    };
    \vertex [above=0.15em of w] {\(\mathbf{k},n,m,\omega\)};
    \vertex [below=0.15em of w0] {\(\mathbf{k}',n',m',\omega'\)};
    \end{feynman}
    \end{tikzpicture},
\end{align}
\end{subequations}
where $\aveHarm{\bullet}$ denotes the expectation as in \Eref{def_ave} but with $\action$ replaced solely by $\actionHarm$.

In \Eref{bare_propagator_feynman}, we have introduced the Feynman diagram for a bare propagator, to be read from right to left and parameterised by generally eight, but due to the $\delta$-functions effectively only four, variables: 
The momentum $\kvec$, in lieu of position $\xvec$ as in \Eref{def_Fourier}, the $\phi$-mode $n$, in lieu of the orientation $\phi$, the $\psi$-mode $m$, in lieu of the reorientation angle $\psi$ and the 
frequency $\omega$, in lieu of direct time $t$.

From the perturbative part of the action, \Eref{actionPert_Fourier}, we can identify three perturbative vertices, two of which represent the self-propulsion and one representing the tumbling. The self-propulsion vertices are
\begin{subequations}
\label{pertvert1}
\begin{align}
    \begin{tikzpicture}[baseline={([yshift=-0.7ex]current bounding box.center)}]
    \begin{feynman}
    \vertex (a) at (0,0);
    \node[isosceles triangle, isosceles triangle apex angle=60,
    draw,
    rotate=270,
    fill=red,color=red,scale=0.5] (b) at (0.8,0);
    \vertex (c) at (1.6,0);
    \diagram*{
    (a)  -- [red, line width=0.25mm, -] (b) -- [red, line width=0.25mm, -] (c)};
    \vertex [above=0.15em of a] {\(n,m \)};
    \vertex [below=0.15em of c] {\(n',m'\)};
    \end{feynman}
    \end{tikzpicture} &\hat{=} \frac{v}{2\imag} (k_x + \imag k_y)\deltabar({\bf{k}}+{\bf{k}}')\deltabar(\omega+\omega')\delta_{n+1,n'}\delta_{m,m'},\\
    \begin{tikzpicture}[baseline={([yshift=-0.7ex]current bounding box.center)}]
    \begin{feynman}
    \vertex (a) at (0,0);
    \node[isosceles triangle, isosceles triangle apex angle=60,
    draw,
    rotate=90,
    fill=red,color=red,scale=0.5] (b) at (0.8,0);
    \vertex (c) at (1.6,0);
    \diagram*{
    (a)  -- [red, line width=0.25mm, -] (b) -- [red, line width=0.25mm, -] (c)};
    \vertex [above=0.15em of a] {\(n,m \)};
    \vertex [below=0.15em of c] {\(n',m'\)};
    \end{feynman}
    \end{tikzpicture} &\hat{=} \frac{v}{2\imag} (k_x - \imag k_y)\deltabar({\bf{k}}+{\bf{k}}')\deltabar(\omega+\omega')\delta_{n-1,n'}\delta_{m,m'},
\end{align}
\end{subequations}
allowing a shift of the $\phi$-mode up or down by one. The third vertex, the tumble vertex, 
\begin{equation}
\label{pertvert2}
    \begin{tikzpicture}[baseline={([yshift=-0.5ex]current bounding box.center)}]
    \begin{feynman}
    \vertex (a) at (0,0);
    \node[circle,
    fill=red,color=red,scale=0.75] (b) at (0.8,0);
    \vertex (c) at (1.6,0);
    \diagram*{
    (a)  -- [red, line width=0.25mm, -] (b) -- [red, line width=0.25mm, -] (c)};
    \vertex [above=0.15em of a] {\(n,m\)};
    \vertex [below=0.15em of c] {\(n',m'\)};
    \end{feynman}
    \end{tikzpicture} \hat{=} \hspace{0.5em} \gamma \deltabar(\mathbf{k}+\mathbf{k}')\deltabar(\omega+\omega')\delta_{n,n'}\delta_{m+n,m'} \ ,
\end{equation}
counter-intuitively amounts to a shift of the $\psi$-mode $m$ by the $\phi$-mode $n$, rather than the other way around. This is because $m$ determines how $\chi$ is modulated (``oscillates'') with $\psi$. If the $n$th mode in $\phi$, namely $\exp{\imag n \phi}$, experiences a phase-shift by $-\psi$, the $m$th mode in $\psi$ becomes the mode $m-n$. This shift does not change the modulation in $\phi$, so that $n$ remains unchanged.

We are now in a position to calculate observables perturbatively using the bare propagator, \Eref{bareprop}, and perturbative vertices, \Erefs{pertvert1} and \eref{pertvert2}. This is done by expressing any desired observables in terms of expectations under the bilinear action,
\begin{equation}
    \langle\bullet\rangle = \langle\bullet\exp{\actionPert}\rangle_0 = \sum_{n=0}^{\infty}\frac{1}{n!}\langle\bullet \actionPert^n \rangle_0 = \sum_{n=0}^{\infty}\frac{1}{n!}\int\mathcal{D}\chi\mathcal{D}\tilde{\chi}\,\bullet\actionPert^{n}\exp{\actionHarm}.
\end{equation}

In particular, the full propagator $\ave{\chi(\xvec,\phi,\psi,t)\chidagger(\xvec_0,\phi_0,\psi_0,t_0)}$, \Eref{probability_propagator}, can be constructed systematically in this way. Because the bare propagator, \Eref{bareprop}, and perturbative vertices, \Erefs{pertvert1} and \eref{pertvert2}, are time and space translationally invariant, the propagator is so too
\begin{equation}
    \langle \chi_{n,m}(\mathbf{k},\omega)\chidagger_{n',m'}(\mathbf{k}',\omega')\rangle \propto \deltabar(\mathbf{k}+\mathbf{k}')\deltabar(\omega+\omega').
\end{equation}
Though the system displays rotational invariance, the same \textit{cannot} be said for $\phi$ and $\psi$. It is obvious that a shift of $\psi$ changes the dynamics of the particle and hence the propagator entirely. It is more subtle to see how in general $\ave{\chi(\xvec,\phi,\psi,t)\chidagger(\xvec_0,\phi_0,\psi_0,t_0)}$ differs from $\ave{\chi(\xvec,\phi+\alpha,\psi,t)\chidagger(\xvec_0,\phi_0+\alpha,\psi_0,t_0)}$. To explore this, we first employ the rotational invariance of the system as a whole to write
\begin{equation}
\ave{\chi(\xvec,\phi,\psi,t)\chidagger(\xvec_0,\phi_0,\psi_0,t_0)} = \ave{\chi(\xvec_\alpha,\phi+\alpha,\psi,t)\chidagger(\xvec_{0_\alpha},\phi_0+\alpha,\psi_0,t_0)}
\end{equation}
where $\xvec_\alpha$ is $\xvec$ after rotation by $\alpha$
\begin{equation}
    \xvec_\alpha = \left({\begin{array}{cc}
   \cos(\alpha) & -\sin(\alpha)\\
   \sin(\alpha) & \cos(\alpha) \\
  \end{array}}\right) \left({\begin{array}{c}
   x  \\
   y \\
  \end{array}}\right) = \left({\begin{array}{c}
   x\cos(\alpha) - y\sin(\alpha)  \\
   y\cos(\alpha) + x\sin(\alpha) \\
  \end{array}}\right)
\end{equation}
and similarly $\xvec_{0_\alpha}$ is the initial condition under rotation by $\alpha$.
If the propagator $\ave{\chi(\xvec,\phi,\psi,t)\chidagger(\xvec_0,\phi_0,\psi_0,t_0)}$ was invariant under a shift by $\alpha$ of $\phi$ and $\phi_0$ alone, without rotating $\xvec$ and $\xvec_0$, then it would equally be invariant under a rotation of $\xvec$ and $\xvec_0$ alone, without shifting $\phi$ and $\phi_0$. This is clearly not the case, because heading in $\phi_0$, generally the particle's probability to go from $\xvec_0$ to $\xvec$ differs from that of going from $\xvec_{0\alpha}$ to $\xvec_\alpha$.

In the following we calculate observables from the full propagator. All results presented are exact despite the perturbative approach. Such calculations can be performed without having access to the full propagator in closed form by identifying non-zero diagrammatic contributions to the desired observable from the full propagator. We first calculate the mean squared displacement as a function of time averaged over the initial reorientation angle $\psi_0$. We then look to the mean squared displacement as a function of time $t$ for fixed, initial $\psi_0$. Moving to orientation-based observables, we first look at the reorientation angle probability density function, which can be used to demonstrate the idea of transient chirality. We next look at the orientation autocorrelator, which can be used as a measure of orientational persistence. We finally look at position-orientation cross-correlators, and see that the position-perpendicular orientation cross-correlator can be used to identify the direction the particle is \textit{initially} tumbling in (i.e. before the reorientation angle has entirely decorrelated).

\section{Mean-squared displacement}\label{sec:msd}

\subsection{Case 1: Mean Squared Displacement with uniform initial \texorpdfstring{$\psi_0$}{psi0}}\label{sec:msdu}

\subsubsection{Derivation}
The mean squared displacement as a function of time $t$ averaged over initial $\psi_0$ is obtained from
\begin{equation} \label{msdu_eq_re}
    \langle|\mathbf{x}|^2(t)\rangle = \int d^{2}x\,|\xvec|^2\int d\phi d\psi \int d\psi_0\, \frac{1}{2\pi}\langle\chi(\xvec,\phi,\psi,t)\chitilde(\mathbf{0},\phi_0,\psi_0,0)\rangle,
\end{equation}
marginalizing over the final angles $\phi$ and $\psi$ and averaging over the initial reorientation angle $\psi_0$, uniformly distributed with density $1/2\pi$. The MSD ought to be independent of the initial direction $\phi_0$, and hence we could opt to average over this degree of freedom, taking $\int d\phi_0(2\pi)^{-1}$ on the right-hand side. However, we can equally keep $\phi_0$ fixed. Carrying out the integral over $\phi,\psi$ and $\psi_0$ selects their $0-$modes in the transformed representation of the fields, \Eref{def_Fourier}. Expressing further $|\mathbf{x}|^2$ as a  double derivative with respect to $\mathbf{k}$ gives
\begin{equation} \label{msdu_eq}
    \langle|\mathbf{x}|^2(t)\rangle = -\int\mathrm{\dbar}\omega e^{-\imag\omega t} \sum_{n'=-\infty}^{\infty} \eval{\nabla^2_{\mathbf{k}}}_{\mathbf{k}=\mathbf{0}}\langle\chi_{0,0}(\mathbf{k},\omega) \tilde{\chi}_{n',0}(\mathbf{k}',\omega')\rangle e^{-\imag n\phi_0}.
\end{equation}
For this to be equivalent to averaging over $\phi_0$, we would expect the only term in the sum that contributes to be $n'=0$. To see this, we first observe that the Laplacian in $\mathbf{k}$ restricts us to diagrams with at most two self-propulsion vertices as any higher order diagrams in $\mathbf{k}$ would be suppressed upon evaluation at $\mathbf{k}=\mathbf{0}$. In fact, the only non-vanishing contributions are due to diagrams with either exactly one of each vertex or none at all. We demonstrate this by considering an amputated diagram comprised of two identical vertices
\begin{align}
    \begin{tikzpicture}[baseline={([yshift=-0.5ex]current bounding box.center)}]
    \begin{feynman}
    \vertex (a) at (0,0);
    \node[isosceles triangle, isosceles triangle apex angle=60,
    draw,
    rotate=90,
    fill=red,color=red,scale=0.5] (b) at (0.4,0);
    \node[isosceles triangle, isosceles triangle apex angle=60,
    draw,
    rotate=90,
    fill=red,color=red,scale=0.5] (c) at (1.2,0);
    \vertex (d) at (1.6,0);
    \diagram*{
    (a)  -- [red, line width=0.25mm, -] (b) -- [red, line width=0.25mm, -] (c)-- [red, line width=0.25mm, -] (d)};
    \vertex [above=0.15em of a] {\(0,0\)};
    \vertex [below=0.15em of d] {\(-2,0\)};
    \end{feynman}
    \end{tikzpicture} \hat{=} \frac{v^2 (k_x-\imag k_y)^2}{4}G(\mathbf{k},-1,0,\omega)\deltabar^{2}(\mathbf{k}+\mathbf{k}')\deltabar(\omega+\omega'),
\end{align}
so that the incoming propagator has $\phi-$mode $n=-2$ and the outgoing has $n=0$. The $\eval{\nabla^2_{\mathbf{k}}}_{\mathbf{k}=\mathbf{0}}$ clearly must operate on the numerator to even have the potential to produce something non-zero, otherwise we are left with prefactors in $k_x$ (resp. $k_y$) that are killed upon evalutation at $\mathbf{k} = \mathbf{0}$. However $(\partial_{k_x}^{2}+\partial_{k_y}^{2})(k_x \pm \imag k_y)^2=0$, so the only non-zero contribution is due to two different vertices, $(\partial_{k_x}^{2}+\partial_{k_y}^{2})(k_x^2 + k_y^2)=4$. For the vertices to be different and result in $n=0$ on the left, $n'=0$ is needed on the right,
\begin{subequations} \label{selfprop}
\begin{align}
    \begin{tikzpicture}[baseline={([yshift=-0.5ex]current bounding box.center)}]
    \begin{feynman}
    \vertex (a) at (0,0);
    \node[isosceles triangle, isosceles triangle apex angle=60,
    draw,
    rotate=270,
    fill=red,color=red,scale=0.5] (b) at (0.4,0);
    \node[isosceles triangle, isosceles triangle apex angle=60,
    draw,
    rotate=90,
    fill=red,color=red,scale=0.5] (c) at (1.2,0);
    \vertex (d) at (1.6,0);
    \diagram*{
    (a)  -- [red, line width=0.25mm, -] (b) -- [red, line width=0.25mm, -] (c)-- [red, line width=0.25mm, -] (d)};
    \vertex [above=0.15em of a] {\(0,0\)};
    \vertex [below=0.15em of d] {\(0,0\)};
    \end{feynman}
    \end{tikzpicture} \hat{=}& -\frac{v^2 k^2}{4}G(\mathbf{k},1,0,\omega)\deltabar^{2}(\mathbf{k}+\mathbf{k}')\deltabar(\omega+\omega'),\\
    \begin{tikzpicture}[baseline={([yshift=-0.5ex]current bounding box.center)}]
    \begin{feynman}
    \vertex (a) at (0,0);
    \node[isosceles triangle, isosceles triangle apex angle=60,
    draw,
    rotate=90,
    fill=red,color=red,scale=0.5] (b) at (0.4,0);
    \node[isosceles triangle, isosceles triangle apex angle=60,
    draw,
    rotate=270,
    fill=red,color=red,scale=0.5] (c) at (1.2,0);
    \vertex (d) at (1.6,0);
    \diagram*{
    (a)  -- [red, line width=0.25mm, -] (b) -- [red, line width=0.25mm, -] (c)-- [red, line width=0.25mm, -] (d)};
    \vertex [above=0.15em of a] {\(0,0\)};
    \vertex [below=0.15em of d] {\(0,0\)};
    \end{feynman}
    \end{tikzpicture} \hat{=}& -\frac{v^2 k^2}{4}G(\mathbf{k},-1,0,\omega)\deltabar^{2}(\mathbf{k}+\mathbf{k}')\deltabar(\omega+\omega').
\end{align}
\end{subequations}

In addition to these two diagrams, there are those with no self-propulsion vertices as well as, in principle, those with a single self-propulsion vertex. However, in these single vertex diagrams, the Laplacian necessarily differentiates the bare propagator, \Eref{bareprop}, once, which is even in $\mathbf{k}$ and thus vanishes under this operation. All contributing diagrams thus have the same incoming and outgoing $\phi-$mode $n=0$, which corresponds to integrating over the final $\phi$ and initial $\phi_0$. We further observe that any number of tumbles outside of this vertex pair will contribute, as each vertex will cause a `decrease' of the $\psi-$mode $m$ by $0$, see \Eref{pertvert2}. As $n=\pm1$ in the intermediate legs in \Eref{selfprop}, no tumble can take place here. Motivated by the need of a diagram capturing an arbitrary number of tumbles with $n=0$ and the same $m$ at both ends we introduce
\begin{equation}\label{gluon_diagrams}
\begin{split}
    \begin{tikzpicture}[baseline={([yshift=-0.5ex]current bounding box.center)}]
    \begin{feynman}
    \vertex (a) at (0,0);
    \vertex (b) at (1.6,0);
    \diagram*{
    (a)  -- [red, line width=0.25mm, gluon] (b)};
    \vertex [above=0.15em of a] {\(0,m\)};
    \vertex [below=0.15em of b] {\(0,m\)};
    \end{feynman}
    \end{tikzpicture} =
    \begin{tikzpicture}[baseline={([yshift=-0.5ex]current bounding box.center)}]
    \begin{feynman}
    \vertex (a) at (0,0);
    \vertex (c) at (0.8,0);
    \diagram*{
    (a)  -- [red, line width=0.25mm, -] (c)};
    \vertex [above=0.15em of a] {\(0,m\)};
    \vertex [below=0.15em of c] {\(0,m\)};
    \end{feynman}
    \end{tikzpicture}+
    \begin{tikzpicture}[baseline={([yshift=-0.5ex]current bounding box.center)}]
    \begin{feynman}
    \vertex (a) at (0,0);
    \node[circle,
    fill=red,color=red,scale=0.75] (b) at (0.8,0);
    \vertex (c) at (1.6,0);
    \diagram*{
    (a)  -- [red, line width=0.25mm, -] (b) -- [red, line width=0.25mm, -] (c)};
    \vertex [above=0.15em of a] {\(0,m\)};
    \vertex [below=0.15em of c] {\(0,m\)};
    \end{feynman}
    \end{tikzpicture} +
    \begin{tikzpicture}[baseline={([yshift=-0.5ex]current bounding box.center)}]
    \begin{feynman}
    \vertex (a) at (0,0);
    \node[circle,
    fill=red,color=red,scale=0.75] (b) at (0.8,0);
    \node[circle,
    fill=red,color=red,scale=0.75] (b) at (1.6,0);
    \vertex (d) at (2.4,0);
    \diagram*{
    (a)  -- [red, line width=0.25mm, -] (b) -- [red, line width=0.25mm, -] (c) -- [red, line width=0.25mm, -] (d)};
    \vertex [above=0.15em of a] {\(0,m\)};
    \vertex [below=0.15em of d] {\(0,m\)};
    \end{feynman}
    \end{tikzpicture}+ \ldots
\end{split}
\end{equation}
where the diagrams on the right correspond to $0,1,2\ldots$ tumbles respectively. Inspecting these one by one we find
\begin{subequations}\label{tumblesums}
\begin{align}
    \begin{tikzpicture}[baseline={([yshift=-0.5ex]current bounding box.center)}]
    \begin{feynman}
    \vertex (a) at (0,0);
    \vertex (c) at (0.8,0);
    \diagram*{
    (a)  -- [red, line width=0.25mm, -] (c)};
    \vertex [above=0.15em of a] {\(0,m\)};
    \vertex [below=0.15em of c] {\(0,m\)};
    \end{feynman}
    \end{tikzpicture} \hspace{0.5em} \hat{=}& \hspace{0.5em} G(\mathbf{k},0,0,\omega)\deltabar^{2}(\mathbf{k}+\mathbf{k'})\deltabar(\omega+\omega'),\\
    \begin{tikzpicture}[baseline={([yshift=-0.5ex]current bounding box.center)}]
    \begin{feynman}
    \vertex (a) at (0,0);
    \node[circle,
    fill=red,color=red,scale=0.75] (b) at (0.8,0);
    \vertex (c) at (1.6,0);
    \diagram*{
    (a)  -- [red, line width=0.25mm, -] (b) -- [red, line width=0.25mm, -] (c)};
    \vertex [above=0.15em of a] {\(0,m\)};
    \vertex [below=0.15em of c] {\(0,m\)};
    \end{feynman}
    \end{tikzpicture} \hspace{0.5em} \hat{=}& \hspace{0.5em} G(\mathbf{k},0,0,\omega)\gamma G(\mathbf{k},0,0,\omega)\deltabar^{2}(\mathbf{k}+\mathbf{k}')\deltabar(\omega+\omega'),\\
   \begin{tikzpicture}[baseline={([yshift=-0.5ex]current bounding box.center)}]
    \begin{feynman}
    \vertex (a) at (0,0);
    \node[circle,
    fill=red,color=red,scale=0.75] (b) at (0.8,0);
    \node[circle,
    fill=red,color=red,scale=0.75] (b) at (1.6,0);
    \vertex (d) at (2.4,0);
    \diagram*{
    (a)  -- [red, line width=0.25mm, -] (b) -- [red, line width=0.25mm, -] (c) -- [red, line width=0.25mm, -] (d)};
    \vertex [above=0.15em of a] {\(0,m\)};
    \vertex [below=0.15em of d] {\(0,m\)};
    \end{feynman}
    \end{tikzpicture} \hspace{0.5em} \hat{=}& \hspace{0.5em} G(\mathbf{k},0,0,\omega)\left[\gamma G(\mathbf{k},0,0,\omega)\right]^{2}\deltabar^{2}(\mathbf{k}+\mathbf{k}')\deltabar(\omega+\omega'),
\end{align}
\end{subequations}
which can be summed as a geometric series,
\begin{align}\label{gluon1}
    \begin{tikzpicture}[baseline={([yshift=-0.5ex]current bounding box.center)}]
    \begin{feynman}
    \vertex (a) at (0,0);
    \vertex (b) at (1.6,0);
    \diagram*{
    (a)  -- [red, line width=0.25mm, gluon] (b)};
    \vertex [above=0.15em of a] {\(0,m\)};
    \vertex [below=0.15em of b] {\(0,m\)};
    \end{feynman}
    \end{tikzpicture}
    \hspace{0.5em}\hat{=}&\hspace{0.5em} G(\mathbf{k},0,0,\omega)\deltabar^{2}(\mathbf{k}+\mathbf{k}')\deltabar(\omega+\omega')\sum_{i=0}^{\infty}\left[\gamma G(\mathbf{k},0,0,\omega)\right]^{i}\nonumber\\
    =&\hspace{0.5em} \frac{G(\mathbf{k},0,0,\omega)\deltabar^{2}(\mathbf{k}+\mathbf{k}')\deltabar(\omega+\omega')}{1-\gamma G(\mathbf{k},0,0,\omega)}.
\end{align}
Using the definition of $G$ in \Eref{bareprop}, we thus arrive at 
\begin{align}\label{gluon}
    \begin{tikzpicture}[baseline={([yshift=-0.5ex]current bounding box.center)}]
    \begin{feynman}
    \vertex (a) at (0,0);
    \vertex (b) at (1.6,0);
    \diagram*{
    (a)  -- [red, line width=0.25mm, gluon] (b)};
    \vertex [above=0.15em of a] {\(0,m\)};
    \vertex [below=0.15em of b] {\(0,m\)};
    \end{feynman}
    \end{tikzpicture}
    \hspace{0.5em}\hat{=}&\frac{\deltabar^{2}(\mathbf{k}+\mathbf{k}')\deltabar(\omega+\omega')}{-\imag\omega +D_x k^2 +\gamma - \gamma + r}\nonumber\\=&\hspace{0.5em} \eval{G(\mathbf{k},0,0,\omega)}_{\gamma = 0}\deltabar^{2}(\mathbf{k}+\mathbf{k}')\deltabar(\omega+\omega'),
\end{align}
where $\eval{G(\mathbf{k},0,0,\omega)}_{\gamma=0}$ is in fact the propagator for a purely diffusive particle in Fourier space. The MSD with uniform initial $\psi_0$ can thus be written as a sum of three contributions: one without self-propulsion vertices and any number of tumbles, and two with two self-propulsion vertices proceeded and followed by any number of tumbles:
\begin{align}\label{msdu}
\langle|\mathbf{x}|^2(t)\rangle &= -\int \dbar \omega\, e^{-i\omega t} \nabla_{\bf{k}}^{2}\biggl[\begin{tikzpicture}[baseline={([yshift=-0.7ex]current bounding box.center)}]
    \begin{feynman}
    \vertex (w) at (0,0);
    \vertex (w0) at (1.2,0);
    \diagram* {(w) -- [red, line width=0.25mm, gluon] (w0),
    };
    \vertex [above=0.15em of w] {\(0,0\)};
    \vertex [below=0.15em of w0] {\(0,0\)};
    \end{feynman}
    \end{tikzpicture} + \begin{tikzpicture}[baseline={([yshift=-0.5ex]current bounding box.center)}]
    \begin{feynman}
    \vertex (a) at (0,0);
    \node[isosceles triangle, isosceles triangle apex angle=60,
    draw,
    rotate=90,
    fill=red,color=red,scale=0.5] (b) at (0.8,0);
    \node[isosceles triangle, isosceles triangle apex angle=60,
    draw,
    rotate=270,
    fill=red,color=red,scale=0.5] (c) at (1.6,0);
    \vertex (d) at (2.4,0);
    \diagram*{
    (a)  -- [red, line width=0.25mm, gluon] (b) -- [red, line width=0.25mm, -] (c)-- [red, line width=0.25mm, gluon] (d)};
    \vertex [above=0.15em of a] {\(0,0\)};
    \vertex [below=0.15em of d] {\(0,0\)};
    \end{feynman}
    \end{tikzpicture} + \begin{tikzpicture}[baseline={([yshift=-0.5ex]current bounding box.center)}]
    \begin{feynman}
    \vertex (a) at (0,0);
    \node[isosceles triangle, isosceles triangle apex angle=60,
    draw,
    rotate=270,
    fill=red,color=red,scale=0.5] (b) at (0.8,0);
    \node[isosceles triangle, isosceles triangle apex angle=60,
    draw,
    rotate=90,
    fill=red,color=red,scale=0.5] (c) at (1.6,0);
    \vertex (d) at (2.4,0);
    \diagram*{
    (a)  -- [red, line width=0.25mm, gluon] (b) -- [red, line width=0.25mm, -] (c)-- [red, line width=0.25mm, gluon] (d)};
    \vertex [above=0.15em of a] {\(0,0\)};
    \vertex [below=0.15em of d] {\(0,0\)};
    \end{feynman}
    \end{tikzpicture}\biggr]_{{\bf{k} }={\bf{0}}},\\
&= \int\mathrm{\dbar}\omega e^{-\imag\omega t} \biggl[\frac{4D_{x}}{(-\imag\omega + r)^2} + \frac{2v^{2}}{(-\imag\omega + r)^{2}(-\imag\omega + D_\phi + \gamma + r)}\biggr].
\end{align}
Performing the integral and taking $r\to 0$, we find
\begin{align}\label{msd_uniform}
\langle|\mathbf{x}|^2(t)\rangle &= 4D_{x}t + \frac{2v^{2}}{(D_\phi + \gamma)^{2}}\bigl[e^{-(D_\phi + \gamma)t} - 1 + (D_\phi + \gamma)t\bigr].
\end{align}
The MSD takes the form of that of an ABP that experiences tumbles \cite{Bechinger2016,CatesTailleur:2013}. As discussed below, see \Sref{sec:analysis}, for large $t$ we can read off the effective diffusive constant $D_{\textrm{eff}}$, as $\lim_{t\to\infty}\langle |\mathbf{x}|^2(t)\rangle/4t$, which produces $D_{\textrm{eff}} = D_x + v^2/(2(D_\phi + \gamma))$, with rotational diffusion and tumbling appearing on the same footing. At small $t$, the square bracket picks up quadratically in $t$, indicating ballistic motion until rotational diffusion or tumbling renders the motion diffusive. This observable does not capture the difference between rotational diffusion and tumbling, as $D_\phi$ and $\gamma$ appear everywhere only as their sum. This changes once we allow the initial increment $\psi_0$ to be fixed, as discussed next.

\subsubsection{Analysis}\label{sec:analysis} 

Here, we characterize the dynamics of the particle in the present case by analysing \Eref{msd_uniform}. We assume that tumbling dominates over orientational diffusion, $D_\phi \ll \gamma$, so that $\tau_p \approx \tau_t$. When considering a uniform initial reorientation angle, the MSD is independent of the correlation time of the reorientation angle $\tau_\psi = 1/D_\psi$. This results in only two time regimes of interest, namely the short-time regime, where $t\ll\tau_p = 1/(D_\phi + \gamma$), and the long-time regime, where $t\gg\tau_p$.

\begin{figure}[b!]
    \centering
    \includegraphics[width=\textwidth]{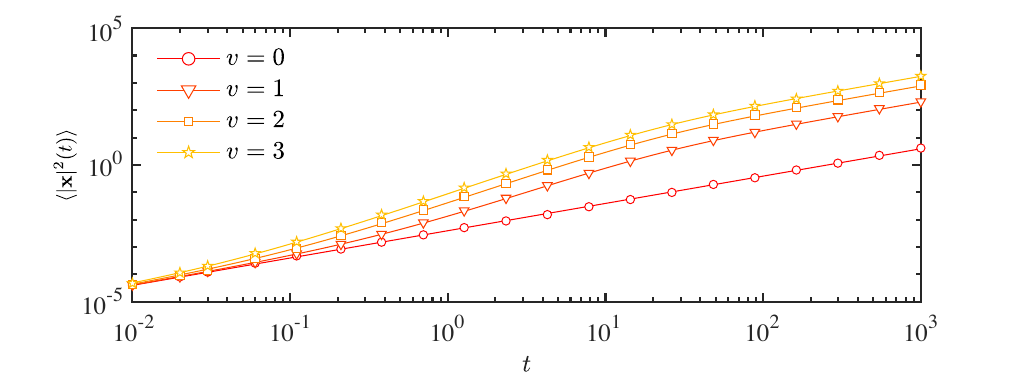}
    \caption{\textit{Mean squared displacement with uniformly chosen initial $\psi_0$} --- The MSD, \Eref{msd_uniform}, with uniform initial $\psi_0$ for $\tau_p = 1$, $D_{x} = 0.01$ and a range of self-propulsion velocities $v$, as indicated. Symbols indicate numerical simulations, solid lines \Eref{msd_uniform}.}
    \label{fig:difftraj}
\end{figure}

\paragraph{Short-time Regime 
\texorpdfstring{$t \ll \tau_p$}{t<<taup}:}
\label{sec:short}
In the short-time regime $t\ll\tau_p$, the orientation angle has not been considerably acted on by either the rotational noise or tumbles. To demonstrate this, we expand the MSD, \Eref{msd_uniform}, to 
\begin{equation}\label{short}
    \langle|\mathbf{x}|^2(t)\rangle = 4D_{x}t + v^2t^2 + \mathcal{O}(t^3),
\end{equation}
The particle's dynamics is clearly dominated by the translational diffusion, $4D_{x}t$, before the ballistic motion takes over, $v^2 t^2$, as long as the particle's orientation $\phi$ has not decorrelated.

\paragraph{Long-time Regime \texorpdfstring{$t \gg \tau_p$}{t>>taup}:}
\label{sec:long}
In the long-time regime $t \gg \tau_p$, the particle's dynamics is bound to be diffusive with effective diffusivity $D_\text{eff}$. Assuming an MSD of the form $4D_\text{eff}t$, we can determine $D_\text{eff}$ from \Eref{msd_uniform} as 
\begin{align}\label{long}
    D_{\text{eff}}  = \lim_{t\to\infty}\frac{\langle|\mathbf{x}|^2(t)\rangle}{4t} = \biggl[D_{x} + \frac{v^2}{2(D_\phi + \gamma)}\biggr].
\end{align}

\subsection{Case 2: Mean Squared Displacement with fixed initial \texorpdfstring{$\psi_0$}{psi0}} \label{sec:msdf}

\subsubsection{Derivation}
The MSD for a fixed initial reorientation angle $\psi_0$ in two dimensions
\begin{equation}\label{msdf_equation}
    \langle|\mathbf{x}|^2(\psi_0, t)\rangle = -\int d\omega e^{-\imag\omega t} \sum_{m'=-\infty}^{\infty} \eval{\nabla^{2}_{\mathbf{k}}}_{\mathbf{k}=0}\langle \chi_{0,0} (\mathbf{k},\omega) \tilde{\chi}_{0,m'}(\mathbf{k}',\omega')\rangle e^{-im' \psi_0},
\end{equation}
which can be thought of as being obtained by averaging over the initial orientation $\phi_0$, thus relieving us of the need to argue about the initial $\phi-$mode being $n'=0$. By fixing $\psi_0$, and equivalently not averaging over this variable, we now instead sum over its Fourier modes. \Eref{msdf_equation} contains the MSD for uniform initial $\psi_0$ as the contribution for $m'=0$, \Eref{msd_uniform}, plus corrections from $m'\neq 0$. Finding the MSD, \Eref{msdf_equation}, for fixed initial $\psi_0$ is a case of identifying relevant diagrams for each $m'$. Starting with $\psi-$mode $m'\neq 0$ on the right, the tumble vertex \Eref{pertvert2} needs to reduce this to $m=0$ on the left using $n\neq 0$. The only mechanism generating $n\neq 0$ is due to the self-propulsion vertices, \Eref{pertvert1}. Because the Laplacian removes higher orders, there can be at most two self-propulsion vertices, as discussed above, taking the $\phi-$mode from $n'=0$ on the right to $n=0$ on the left. The order of the self-propulsion vertices is determined by the sign of $m'$; $m'>0$ on the right, we need to $n=1$ so that each subsequent tumble decreases $m$ by 1, but for $m'<0$ we need $n=-1$
\begin{multline}
    \langle \chi_{0,0}({\bf{k}},\omega)\tilde{\chi}_{0,m'}({\bf{k}}',\omega')\rangle \\ \simeq
    \begin{cases}\begin{tikzpicture}[baseline={([yshift=0.6ex]current bounding box.center)}]
    \begin{feynman}
    \vertex (a) at (-0.15,0);
    \node[isosceles triangle, isosceles triangle apex angle=60,
    draw,
    rotate=270,
    fill=red,color=red,scale=0.5] (a1) at (0.4,0);
    \node[circle,
    fill=red,color=red,scale=0.5] (b) at (0.8,0);
    \vertex (c) at (1.2,0);
    \vertex (d) at (1.6,0);
    \node[circle,
    fill=red,color=red,scale=0.5] (e) at (2.0,0);
    \node[isosceles triangle, isosceles triangle apex angle=60,
    draw,
    rotate=90,
    fill=red,color=red,scale=0.5] (e1) at (2.4,0);
    \vertex (f) at (2.95,0);
    \vertex (A) at (0.5,-0.1);
    \vertex (B) at (2.3,-0.1);
    \diagram*{
    (a)  -- [red, line width=0.25mm, gluon] (a1) -- [red, line width=0.25mm, -] (c)-- [red, line width=0.25mm, dotted] (d)-- [red, line width=0.25mm, -] (e1)-- [red, line width=0.25mm, gluon] (f)};
    \vertex [above=0.15em of a] {\(0,0\)};
    \vertex [below=0.15em of f] {\(0,m'\)};
    \draw [thick, black,decorate,decoration={brace,amplitude=12pt,mirror}](A) -- (B) node[black,midway,yshift=-0.6cm]{\footnotesize $m'$};
    \end{feynman}
    \end{tikzpicture}\hat{=}\hspace{0.5em} \frac{-v^{2}k^{2}}{4}\biggl( \eval{G(\mathbf{k},0,m',\omega)}_{\gamma=0} \eval{G(\mathbf{k},0,0,\omega)}_{\gamma = 0}\gamma^{m'} \prod_{j=0}^{m'} G(\mathbf{k},1,j,\omega)\biggr) & m'>0 \\ \begin{tikzpicture}[baseline={([yshift=0.6ex]current bounding box.center)}]
    \begin{feynman}
    \vertex (a) at (-0.15,0);
    \node[isosceles triangle, isosceles triangle apex angle=60,
    draw,
    rotate=90,
    fill=red,color=red,scale=0.5] (a1) at (0.4,0);
    \node[circle,
    fill=red,color=red,scale=0.5] (b) at (0.8,0);
    \vertex (c) at (1.2,0);
    \vertex (d) at (1.6,0);
    \node[circle,
    fill=red,color=red,scale=0.5] (e) at (2.0,0);
    \node[isosceles triangle, isosceles triangle apex angle=60,
    draw,
    rotate=270,
    fill=red,color=red,scale=0.5] (e1) at (2.4,0);
    \vertex (f) at (2.95,0);
    \vertex (A) at (0.5,-0.1);
    \vertex (B) at (2.3,-0.1);
    \diagram*{
    (a)  -- [red, line width=0.25mm, gluon] (a1) -- [red, line width=0.25mm, -] (c)-- [red, line width=0.25mm, dotted] (d)-- [red, line width=0.25mm, -] (e1)-- [red, line width=0.25mm, gluon] (f)};
    \vertex [above=0.15em of a] {\(0,0\)};
    \vertex [below=0.15em of f] {\(0,m'\)};
    \draw [thick, black,decorate,decoration={brace,amplitude=12pt,mirror}](A) -- (B) node[black,midway,yshift=-0.6cm]{\footnotesize $-m'$};
    \end{feynman}
    \end{tikzpicture}\hat{=}\hspace{0.5em} \frac{-v^{2}k^{2}}{4}\biggl(\eval{G(\mathbf{k},0,m',\omega)}_{\gamma=0}\eval{G(\mathbf{k},0,0,\omega)}_{\gamma=0}\gamma^{-m'} \prod_{j=0}^{-m'} G(\mathbf{k},-1,j,\omega)\biggr) & m'<0\end{cases}
\end{multline}

Identifying equal terms and using the definition of the bare propagator, \Eref{bareprop}, the MSD, \Eref{msdf_equation}, for fixed initial $\psi_0$ can be expressed in terms of \Eref{msd_uniform}, the MSD for uniform initial $\psi_0$ corresponding to $m'=0$, and a correction term
\begin{align}\label{msdf_int}
    \langle|\mathbf{x}|^2(\psi_0, t)\rangle = \langle|\mathbf{x}|^2(t)\rangle +2v^{2}\int\mathrm{\dbar}\omega e^{-\imag\omega t} \sum_{m=1}^{\infty}\cos{(m\psi_0)}\biggl[\frac{\gamma^{m}}{(-\imag\omega + r)(-\imag\omega + D_\psi m^2 + r)}&\nonumber\\ 
    \cross\prod_{j=0}^{m}\frac{1}{-\imag\omega + D_\phi + D_\psi j^2 +\gamma +r}&\biggr].
\end{align}
where we have used $\cos{(m\psi_0)}=(e^{\imag m \psi_0}+e^{-\imag m \psi_0})/2$ to reduce the sum in \Eref{msdf_equation} to one over positive $m$ only. Integrating and taking $r \to 0$ we find
\begin{multline}\label{msdf}
    \langle|\mathbf{x}|^2(\psi_0, t)\rangle = \langle|\mathbf{x}|^2(t)\rangle + 2v^2\sum_{m=1}^{\infty} \cos{(m\psi_0)} \gamma^{m}\biggl[\frac{1}{D_{\psi}m^2}\prod_{j=0}^{m}\frac{1}{D_\phi +D_\psi j^2 + \gamma}\\
    \fl-\frac{e^{-D_\psi m^2 t}}{D_{\psi}m^2}\prod_{j=0}^{m}\frac{1}{D_\phi +D_\psi (j^2 - m^2) + \gamma}\\ 
    +\sum_{\ell=0}^{m} \frac{e^{-(D_\phi + D_\psi \ell^2 + \gamma)t}}{(D_\phi + D_\psi \ell^2 + \gamma)(D_\phi + D_\psi(\ell^2 - m^2) + \gamma)}\prod^{m}_{j=0, j\neq \ell} \frac{1}{D_\psi (j^2 - \ell^2)} \biggr],
\end{multline}
This concludes the present derivation. As a sanity check for \Eref{msdf} we consider $t\to0$ which should produce $\langle|\mathbf{x}|^2(\psi_0, t\to 0)\rangle = 0$. However, for $t\to 0$, the exponentials leave behind all three terms inside the square bracket, which do not cancel in an obvious way. Yet, by comparing poles in $z$ of rational functions we find
\begin{equation}
    -\prod_{j=0}^{m}\frac{1}{z+j^2} + \prod_{j=0}^{m}\frac{1}{z+j^2-m^2} = \sum_{\ell=0}^{m}\frac{m^2}{(z+\ell^2)(z+\ell^2-m^2)}\prod_{j=0,j\neq\ell}^{m}\frac{1}{j^2-\ell^2},
\end{equation}
which allows \Eref{msdf} to be rewritten as
\begin{multline}\label{msdfv2}
    \langle|\mathbf{x}|^2(\psi_0, t)\rangle = \langle|\mathbf{x}|^2(t)\rangle + 2v^2\sum_{m=1}^{\infty} \cos{(m\psi_0)} \gamma^{m}\biggl[\frac{1-e^{-D_\psi m^2 t}}{D_{\psi}m^2}\prod_{j=0}^{m}\frac{1}{D_\phi +D_\psi (j^2 - m^2) + \gamma}\\ 
    +\sum_{\ell=0}^{m} \frac{e^{-(D_\phi + D_\psi \ell^2 + \gamma)t}-1}{(D_\phi + D_\psi \ell^2 + \gamma)(D_\phi + D_\psi(\ell^2 - m^2) + \gamma)}\prod^{m}_{j=0, j\neq \ell} \frac{1}{D_\psi (j^2 - \ell^2)} \biggr],
\end{multline}
so that all terms in the square bracket vanish as $t\to0$, as does $\langle|\mathbf{x}|^2(t)\rangle$, \Eref{msd_uniform}.

\subsubsection{Analysis}

Here, we show that the MSD suggests a transition from translationally diffusive and subsequently ballistic behavior at short-times to diffusive at long-times, after passing through one of two intermediate regimes that are dependent on how fast the reorientation angle $\psi$ decorrelates. We note that although the MSD for fixed initial reorientation angle $\psi_0$ in \Erefs{msdf} and \eref{msdfv2} still requires multiple sums and products to be carried out, these for large $m$ are strongly suppressed by products of the form $\prod_{j} 1/(j^2 - \ell^2)$, allowing for straight-forward numerical evaluation.

As opposed to the MSD with uniform initial $\psi_0$, \Eref{msd_uniform}, fixed initial $\psi_0$, \Erefs{msdf} and \eref{msdfv2}, reveals the presence of $D_\psi$ as well as a difference in the dependence on $D_\phi$ and $\gamma$, which features as a prefactor $\gamma^m$, rather than only ever as a sum together with $D_\phi$. As one would expect, the effective diffusion constant and short-time ballistic behavior for fixed initial $\psi_0$ is identical to that of uniform initial $\psi_0$. For large $t$, the exponentials in \Eref{msdf} suppress all correction terms  except for a constant, so that $\langle|\mathbf{x}|^2(\psi_0, t)\rangle = \langle|\mathbf{x}|^2(t)\rangle + \mathcal{O}(t^0)$. For small $t$, expanding the exponentials to order $t^2$ shows no contribution to the ballistic or diffusive behavior of the particle compared to uniform initial $\psi_0$, as the director angle $\phi$ is at short times not affected by $\psi_0$ and its distribution. This is discussed in more details below.

For fixed initial $\psi_0$, the dynamics is thus subject to both timescales, the persistence time $\tau_p = 1/(D_\phi + \gamma)$ and the time $\tau_\psi = 1/D_\psi$ for the reorientation angle to decorrelate. As in the case of uniform $\psi_0$, at short times the particle is subject to few or no tumbles by angle $\psi$, and therefore $\tau_\psi$ is not `visible' in the trajectory. At very long times, both $\phi$ and $\psi$ completely decorrelate, displaying the long-time behavior of an active Brownian particle with run-and-tumble dynamics. However, there are two distinct intermediate regimes, corresponding to the cases where the decorrelation of $\psi$ is faster or slower than that of $\phi$, to be discussed in the following:

\paragraph{Limit of fast decorrelation of reorientation angle \texorpdfstring{$1/D_\psi = \tau_\psi \ll t \ll \tau_p = 1/(D_\phi+\gamma)$}{1/Dpsi=taupsi>>t>>taup=1/(Dphi+gamma)}:}
\label{sec:scenario1}

If the reorientation angle $\psi$ decorrelates faster than the director $\phi$, we expect the dynamics to be independent of the initial reorientation angle, because few tumbles occur before the reorientation angle decorrelates. Recognising that $m$ in  \Erefs{msdf} and \eref{msdfv2} accounts for the number of tumbles, we can truncate the sums at $m=1$. In this regime we find that
\begin{equation}\label{fast}
    \langle|\mathbf{x}|^2(\psi_0,t)\rangle \simeq 4D_{x}t +v^2t^2,
\end{equation}
the term linear in $\cos{(\psi_0)}$ is $\mathcal{O}(t^3)$, so that the MSD is given by \Eref{short}. This is consistent with $\psi_0$ having no discernible effect on the particle's trajectory within the persistence time of the particle.

\paragraph{Limit of slow decorrelation of reorientation angle \texorpdfstring{$\tau_p \ll t \ll \tau_\psi$}{taup>>t>>taupsi}:}
\label{sec:scenario2}
\begin{figure}[t!]
    \centering
    \includegraphics[width=\textwidth]{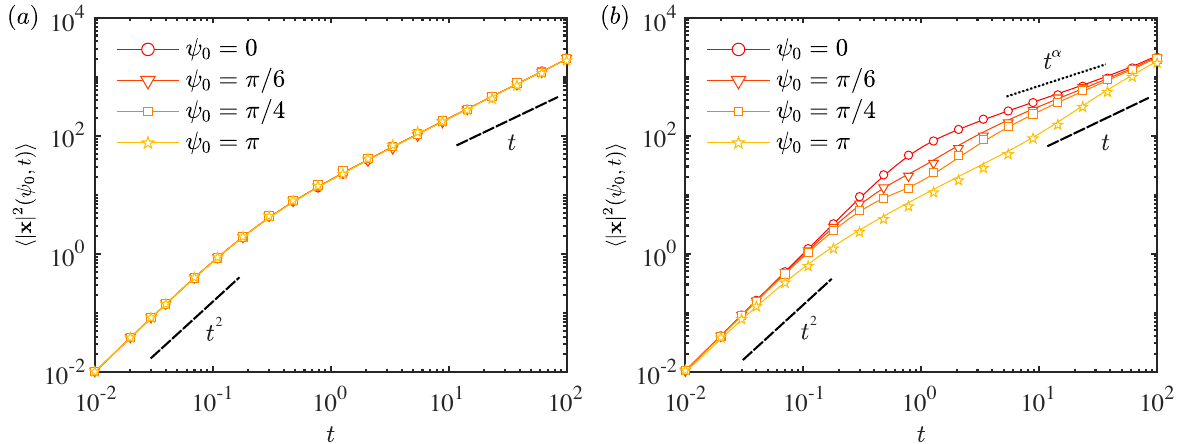}
    \caption{Mean squared displacement with fixed initial $\psi_0$ --- The MSD for a TCAP with fixed initial $\psi_0$, \Erefs{msdf} and \eref{msdfv2}, is plotted in the limit of slow decorrelation of $\psi$ with (a) low tumble rate and high diffusivity of the director $\gamma = 0.1$, $D_\phi=10$, $D_\psi=0.1$, $D_{x} = 0.01$ and (b) high tumble rate and low diffusivity of the director $\gamma = 10$, $D_\phi=0.1$, $D_\psi=0.1$, $D_{x} = 0.01$ for various values of $\psi_0$. Symbols indicate numerical simulations, solid lines \Eref{msdf}. The dashed lines indicate ballistic behaviour $\propto t^2$ and diffusive behaviour linear in $t$ respectively. The dotted line indicates subdiffusive behaviour $\propto t^\alpha$ with $\alpha<1$.}\label{fig:intermediate}
\end{figure}
In this regime $\psi$ decorrelates slower than $\phi$. We expect the MSD to take the form of the long-time MSD with corrections that depend on $\psi_0$, i.e.
\begin{equation}\label{slow}
    \langle|\mathbf{x}|^2(\psi_0,t)\rangle \simeq 4D_\text{eff}t + ... 
\end{equation}
If the diffusion, $D_\phi$, of the director is small, so that the persistence time is due to the tumbling, a small reorientation angle $\psi_0$ will help maintain essentially ballistic motion. For large $D_\phi$, the dependence on $\psi_0$ will be weaker. This is illustrated in \Fref{fig:intermediate}, showing virtually no dependence on $\psi_0$ if $\gamma$ is small and $D_\phi$ is high. Surprisingly, for small $\psi_0$, the MSD shows intermediate, subdiffusive behaviour. For large $\gamma \gg D_\phi,D_\psi$ the sums in \Erefs{msdf} and \eref{msdfv2} can no longer be truncated at small $m$. To capture the detailed dependence on $\psi_0$ the sum has to be carried out to large $m\sim\gamma T$, the expected number of tumbles. The data shown in \Eref{fig:intermediate} is plotted up to $m=300$.

\section{Reorientation angle probability density function}\label{sec:marginal}
\begin{figure}[t!]
\centering
\includegraphics[width=\textwidth]{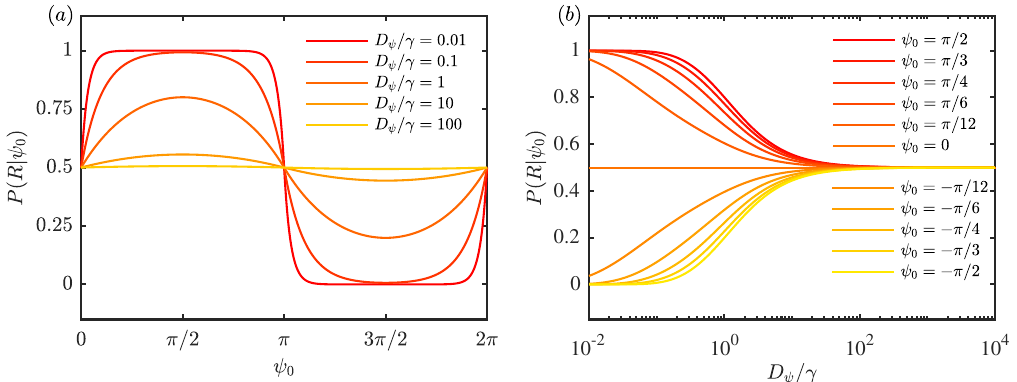}
\caption{Emergent transient chirality --- The probability $P(R|\psi_0)$, \Eref{eq:transientchiralityevidence}, to observe a right-handed tumble following a tumble with reorientation $\psi_0$ as (a) a function of $\psi_0$ for a range of $D_\psi/\gamma$ and (b) a function of $D_\psi/\gamma$ for a range of $\psi_0$.}
\label{fig:transientchirality}
\end{figure}

To illustrate transient chirality, we consider the dynamics of the reorientation angle. We denote $P(\psi,t | \psi_0)$ the probability density that the reorientation angle is equal to $\psi$ at time $t$ knowing that it was initially, i.e. at time $t=0$, at $\psi_0$. The reorientation angle diffuses over a ring of length $2 \pi$, with periodic boundary conditions $P(\psi + 2n \pi, t | \psi_0) = P(\psi, t | \psi_0)$. We can find $P(\psi,t|\psi_0)$ by marginalising the full propagator
\begin{subequations}
\begin{align}
    P(\psi,t|\psi_0) &= \int d^{2}x\,\int d\phi d\phi_0\,\langle\chi(\xvec,\phi,\psi,t)\chitilde(\mathbf{0},\phi_0,\psi_0,0)\rangle,\\
    &= \frac{1}{2\pi}\int\mathrm{\dbar}\omega e^{-\imag\omega t} \sum_{m,m'=-\infty}^{\infty} \biggl[\langle\chi_{0,m}(\mathbf{k},\omega) \tilde{\chi}_{0,m'}(\mathbf{k}',\omega')\rangle e^{\imag(m\psi- m'\psi_0)}\biggr]_{\mathbf{k}=\mathbf{0}}.
\end{align}
\end{subequations}
Fixing both $\psi$ and $\psi_0$ results in a double Fourier sum over the $\psi-$modes. Evaluating at $\mathbf{k}=\mathbf{0}$ restricts us to diagrams with no self-propulsion vertices as these introduce prefactors in $k_x$ and $k_y$, \Eref{pertvert1}. This means all contributing diagrams maintain $\phi-$mode $n=0$, and thus any number of tumble vertices, \Eref{pertvert2}, can contribute as they have no effect on the $\psi-$mode, resulting in $m=m'$, \Eref{gluon_diagrams}. As a result, we arrive at
\begin{subequations}
\begin{align}
    P(\psi,t|\psi_0) &= \frac{1}{2\pi}\int \dbar \omega\, e^{-i\omega t} \sum_{m=-\infty}^{\infty}\biggl[e^{\imag m(\psi-\psi_0)}\begin{tikzpicture}[baseline={([yshift=-0.7ex]current bounding box.center)}]
    \begin{feynman}
    \vertex (w) at (0,0);
    \vertex (w0) at (1.6,0);
    \diagram* {(w) -- [red, line width=0.25mm, gluon] (w0),
    };
    \vertex [above=0.15em of w] {\(0,m\)};
    \vertex [below=0.15em of w0] {\(0,m\)};
    \end{feynman}\end{tikzpicture}\biggr]_{\mathbf{k}=\mathbf{0}}\\
    &= \frac{1}{2\pi}\int \dbar \omega\, e^{-i\omega t} \sum_{m=-\infty}^{\infty}\biggl[\frac{e^{\imag m(\psi-\psi_0)}}{-\imag \omega + D_\psi m^2 + r}\biggr].
\end{align}
\end{subequations}
Integrating and taking $r\to 0$, we arrive at
\begin{subequations}
\begin{align}\label{marginal}
P(\psi,t | \psi_0) &= \frac{1}{2 \pi} \left[1+ 2 \sum_{m=1}^{\infty} \cos\left[m (\psi-\psi_0) \right] e^{-m^2D_{\psi}t}\right] \\
&= \frac{1}{2\pi} \vartheta_3\left(\frac{\psi-\psi_0}{2},e^{-D_{\psi}t}\right),
\end{align}
\label{eq:propjacobi}%
\end{subequations}
the desired result for diffusion on a ring, with $\vartheta_3(z,q)$ the third Jacobi theta function \cite{Gradshteyn:2007}. Given this probability density, we can for instance calculate the probability $P(R|\psi_0)$ that the next tumble will occur with $\psi\in[0,\pi)$, i.e. to the right, having previously tumbled with reorientation angle $\psi_0$, 
\begin{subequations}
\begin{align}\label{right}
P(R | \psi_0) &= \int_{0}^{\infty} d\tau \gamma \exp{-\gamma \tau} \int_0^{\pi} d \psi P(\psi, \tau | \psi_0) \\
&= \frac{1}{2} + \frac{2}{\pi} \sum_{k=0}^{\infty} \frac{\sin[(2k+1)\psi_0]}{2k+1}  \frac{1}{1 + (2k+1)^2 D_\psi/\gamma},
\label{eq:transientchiralityevidence}
\end{align}
\end{subequations}
where $\gamma \exp{-\gamma \tau}$ is the probability density of the next tumble taking place after time $\tau$ has elapsed since the last one.
On the right-hand side of \Eref{eq:transientchiralityevidence}, there remain only two parameters: $\psi_0$ and $D_\psi/\gamma$. As shown in \Fref{fig:transientchirality}, we confirm that for $D_\psi/\gamma \ll 1$, the probability of a right-handed tumble can be particularly large (resp. small) if the reorientation angle at the previous tumble is $\psi_0 \in [0,\pi)$ (resp. $\psi_0 \in [-\pi,0)$).

In the limit where the typical time between tumbles is large compared to the decorrelation time of the reorientation angle, $D_\psi/\gamma \gg 1$, we observe that $P(R | \psi_0)$ becomes asymptotically independent of $\psi_0$. \Fref{fig:transientchirality} justifies the notion of \textit{transient chirality}, as particles consistently tumble in one direction approximately $\gamma/D_{\psi} = \gamma\tau_\psi$ times.

\section{Orientation autocorrelator}\label{sec:auto}
The autocorrelator of the orientation $\mathbf{e}(\phi(t))$, \Eref{def_orr}, for a fixed initial orientation $\phi_0$ is written
\begin{equation}\label{def_orr}
    \langle \mathbf{e}(\phi(t))\cdot\mathbf{e}(\phi_0)\rangle = \langle\cos{(\phi)}\rangle\cos{(\phi_0)}+\langle\sin{(\phi)}\rangle\sin{(\phi_0)}.
\end{equation}
While this can be recognised as the expectation of $\cos{(\phi-\phi_0)}$, we instead calculate $\langle\cos{(\phi)}\rangle$ and $\langle\sin{(\phi)}\rangle$ individually. We first focus on 
\begin{multline}
    \langle\cos{(\phi)}\rangle = \frac{1}{2}\int\dbar\omega e^{-\imag \omega t}\int \dbar^2 k \sum_{n',m'=-\infty}^{\infty} \bigl[\langle \chi_{1,0} (\mathbf{k}=\mathbf{0},\omega) \tilde{\chi}_{n',m'}(\mathbf{k}',\omega')\rangle\\ + \langle \chi_{-1,0} (\mathbf{k}=\mathbf{0},\omega) \tilde{\chi}_{n',m'}(\mathbf{k}',\omega')\rangle\bigr] e^{-\imag n'\phi_0}e^{-\imag m'\psi_0},
\end{multline}
marginalising the probability density $\prob(\xvec,\phi,\psi,t; \xvec_0,\phi_0,\psi_0,t_0)$ in \Eref{probability_propagator} over $\xvec$ and $\psi$, so that $\mathbf{k}=\mathbf{0}$ and $m=0$, and fixing initial $\phi_0$ and $\psi_0$ which results in a sum over Fourier modes for both initial variables $n'$ and $m'$. Euler's $\cos(\phi) = (1/2)(\exp{\imag x} + \exp{-\imag x})$ produces the leading $1/2$ and the two $\phi-$modes $n=1$ and $n=-1$.  Evaluation at $\mathbf{k}=\mathbf{0}$ restricts us to diagrams with no self-propulsion vertices, \Eref{pertvert1}, as these would introduce prefactors of $\mathbf{k}$, so that the $\phi-$mode on the right must match the $\phi-$mode on the left, $n=1$ and $n=-1$ respectively. Any incoming $\psi-$mode $m'$ on the right needs to be reduced to $m=0$ on the left using chains of $m$ tumble vertices, \Eref{pertvert2}. Diagrams with $m'>0$ require $n'=1$ in order to decrease the $\psi-$mode to 0. Similarly, diagrams with $m'<0$ require $n'=-1$. Further, there are bare level contributions from $m'=0$, so that we arrive at
\begin{multline}
   \langle \chi_{1,0} (\mathbf{k},\omega) \tilde{\chi}_{n',m'}(\mathbf{k}',\omega')\rangle + \langle \chi_{-1,0} (\mathbf{k},\omega) \tilde{\chi}_{n',m'}(\mathbf{k}',\omega')\rangle \\ \hat{=}
    \begin{cases}\begin{tikzpicture}[baseline={([yshift=0.6ex]current bounding box.center)}]
    \begin{feynman}
    \vertex (a) at (-0.15,0);
    \vertex (a1) at (0.4,0);
    \node[circle,
    fill=red,color=red,scale=0.5] (b) at (0.8,0);
    \vertex (b1) at (0.8,-0.15);
    \vertex (c) at (1.2,0);
    \vertex (d) at (1.6,0);
    \node[circle,
    fill=red,color=red,scale=0.5] (e) at (2.0,0);
    \vertex (e2) at (2.0, -0.15);
    \vertex (e1) at (2.4,0);
    \vertex (f) at (2.95,0);
    \vertex (A) at (0.5,-0.1);
    \vertex (B) at (2.3,-0.1);
    \diagram*{
    (a1) -- [red, line width=0.25mm, -] (c)-- [red, line width=0.25mm, dotted] (d)-- [red, line width=0.25mm, -] (e1)};
    \vertex [above=0.2em of a1] {\(1,0\)};
    \vertex [above=0.2em of e1] {\(1,m'\)};
    \draw [thick, black,decorate,decoration={brace,amplitude=6pt,mirror}](b1) -- (e2) node[black,midway,yshift=-0.4cm]{\footnotesize $m'$};
    \end{feynman}
    \end{tikzpicture}\hat{=}\hspace{0.5em} \gamma^{m'} \prod_{j=0}^{m'} G(\mathbf{k},-1,j,\omega) & n'=1, m'>0 \\ \begin{tikzpicture}[baseline={([yshift=0.6ex]current bounding box.center)}]
    \begin{feynman}
    \vertex (a) at (0.55,0);
    \vertex (a1) at (0.4,0);
    \node[circle,
    fill=red,color=red,scale=0.5] (b) at (0.8,0);
    \vertex (b1) at (0.8,-0.15);
    \vertex (c) at (1.2,0);
    \vertex (d) at (1.6,0);
    \node[circle,
    fill=red,color=red,scale=0.5] (e) at (2.0,0);
    \vertex (e1) at (2.4,0);
    \vertex (e2) at (2.0, -0.15);
    \vertex (f) at (2.25,0);
    \vertex (A) at (0.5,-0.1);
    \vertex (B) at (2.3,-0.1);
    \diagram*{
    (a1) -- [red, line width=0.25mm, -] (c)-- [red, line width=0.25mm, dotted] (d)-- [red, line width=0.25mm, -] (e1)};
    \vertex [above=0.2em of a] {\(-1,0\)};
    \vertex [above=0.2em of f] {\(-1,m'\)};
    \draw [thick, black,decorate,decoration={brace,amplitude=6pt,mirror}](b1) -- (e2) node[black,midway,yshift=-0.4cm]{\footnotesize $-m'$};
    \end{feynman}
    \end{tikzpicture}\hat{=}\hspace{0.5em} \gamma^{-m'} \prod_{j=0}^{-m'} G(\mathbf{k},1,j,\omega)& n'=-1, m'<0\\
    \begin{tikzpicture}[baseline={([yshift=-0.5ex]current bounding box.center)}]
    \begin{feynman}
    \vertex (a) at (0,0);
    \vertex (c) at (0.8,0);
    \diagram*{
    (a)  -- [red, line width=0.25mm, -] (c)};
    \vertex [above=0.15em of a] {\(1,0\)};
    \vertex [below=0.15em of c] {\(1,0\)};
    \end{feynman}
    \end{tikzpicture} + \begin{tikzpicture}[baseline={([yshift=-0.5ex]current bounding box.center)}]
    \begin{feynman}
    \vertex (a) at (0,0);
    \vertex (c) at (0.8,0);
    \diagram*{
    (a)  -- [red, line width=0.25mm, -] (c)};
    \vertex [above=0.15em of a] {\(-1,0\)};
    \vertex [below=0.15em of c] {\(-1,0\)};
    \end{feynman}
    \end{tikzpicture} \hat{=}\,G(\mathbf{k},1,0,\omega) + G(\mathbf{k},-1,0,\omega) & m'=0\end{cases}
\end{multline}

Simplifying as we did from \Eref{msdf_equation} to \Eref{msdf_int} and using the definition of the bare propagator \Eref{bareprop}, $\langle\cos{(\phi)}\rangle$ can be expressed as 
\begin{multline}
    \langle\cos{(\phi)}\rangle = \frac{1}{2}\int\dbar\omega e^{-\imag \omega t} \biggl(\frac{\cos{(\phi_0)}}{-\imag\omega + D_\phi + \gamma +r}\\ +\sum_{m=1}^{\infty}\cos{(\phi_0+m\psi_0)}\gamma^m \prod_{i=0}^{m}\frac{1}{-\imag\omega+D_\phi+D_\psi i^2+\gamma+r}\biggr).
\end{multline}
Integrating and taking $r \to 0$ we find
\begin{multline}
\langle\cos{(\phi)}\rangle = e^{-(D_\phi+\gamma)t}\cos{(\phi_0)} \\+ \sum_{m=1}^{\infty}\cos{(\phi_0+m\psi_0)}\gamma^m \biggl(\sum_{j=0}^{m}e^{-(D_\phi + D_\psi j^2+\gamma)t}\prod_{i=0,i\neq j}^{m}\frac{1}{D_\psi (i^2 - j^2)}\biggr).
\end{multline}
Similarly, for $\langle\sin{(\phi)}\rangle$ we find
\begin{multline}
\langle\sin{(\phi)}\rangle = e^{-(D_\phi+\gamma)t}\sin{(\phi_0)} \\+ \sum_{m=1}^{\infty}\sin{(\phi_0+m\psi_0)}\gamma^m \biggl(\sum_{j=0}^{m}e^{-(D_\phi + D_\psi j^2+\gamma)t}\prod_{i=0,i\neq j}^{m}\frac{1}{D_\psi (i^2 - j^2)}\biggr),
\end{multline}
\begin{figure}[t!]
\centering
\includegraphics[width=0.6\textwidth]{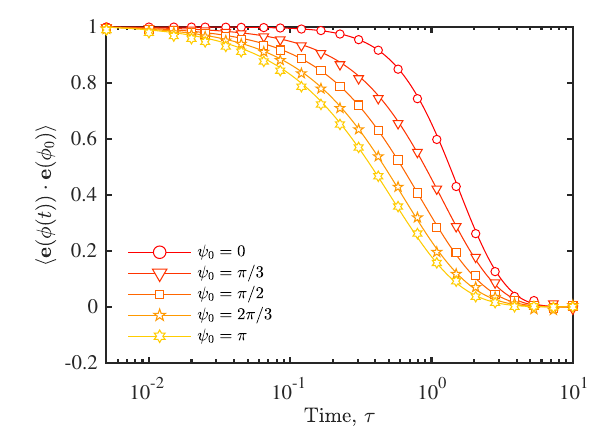}
\caption{Orientation autocorrelator --- $\langle\mathbf{e}(\phi(t))\cdot\mathbf{e}(\phi_0)\rangle$ plotted as a function of time, $t$, for a variety of initial reorientation angles $\psi_0$ with $D_\phi=0, D_\psi=1,\gamma=1$. Symbols indicate numerical simulations, solid lines \Eref{orr_auto}.}
\label{fig:auto}
\end{figure}
so that \Eref{def_orr} gives
\begin{multline}\label{orr_auto}
\langle\mathbf{e}(\phi(t))\cdot\mathbf{e}(\phi_0)\rangle = e^{-(D_\phi+\gamma)t} + \sum_{m=1}^{\infty}\cos{(m\psi_0)}\gamma^m \biggl(\sum_{j=0}^{m}e^{-(D_\phi + D_\psi j^2+\gamma)t}\prod_{i=0,i\neq j}^{m}\frac{1}{D_\psi (i^2 - j^2)}\biggr).
\end{multline}
\Fref{fig:auto} shows the cross-correlator $\langle\mathbf{e}(\phi(t))\cdot\mathbf{e}(\phi_0)\rangle$, \Eref{orr_auto}, for a range of initial $\psi_0$. As it is symmetric under a change of sign of $\psi_0$, we restrict ourselves to $\psi_0\in[0,\pi)$. We observe persistent motion at short times no matter how large $\psi_0$, followed by an exponential decay, with a sharper descent for larger $\psi_0$ as the particle on average tumbles by a larger value. Asymptotically at large $t$, the director decorrelates and the correlator vanishes. In the following section, we use some of the techniques illustrated above to determine the position-orientation cross-correlator.


\section{Position-orientation cross-correlator}\label{sec:cross}
The position-orientation cross-correlator for fixed initial $\psi_0$,
\begin{equation}
    \langle \mathbf{x}(t)\cdot\mathbf{e}(\phi(t))\rangle = \langle x(t)\cos{(\phi(t))\rangle + \langle y(t)\sin{(\phi(t))\rangle}},
\end{equation}
can be derived from the propagator, \Eref{probability_propagator}, via
\begin{multline}\label{xcos}
    \langle x(t)\cos{(\phi(t))}\rangle = \frac{\imag}{2}\int\dbar\omega e^{-\imag \omega t}\sum_{m'=-\infty}^{\infty} \eval{\partial_{k_x}}_{\mathbf{k}=\mathbf{0}}[\langle \chi_{-1,0} (\mathbf{k},\omega) \tilde{\chi}_{0,m'}(\mathbf{k}',\omega')\rangle\\ + \langle \chi_{1,0} (\mathbf{k},\omega) \tilde{\chi}_{0,m'}(\mathbf{k}',\omega')\rangle]e^{-\imag m'\psi_0},
\end{multline}
\begin{multline}\label{ysin}
    \langle y(t)\sin{(\phi(t))}\rangle = \frac{1}{2}\int\dbar\omega e^{-\imag \omega t}\sum_{m'=-\infty}^{\infty} \eval{\partial_{k_y}}_{\mathbf{k}=\mathbf{0}}[\langle \chi_{-1,0} (\mathbf{k},\omega) \tilde{\chi}_{0,m'}(\mathbf{k}',\omega')\rangle\\- \langle \chi_{1,0} (\mathbf{k},\omega) \tilde{\chi}_{0,m'}(\mathbf{k}',\omega')\rangle]e^{-\imag m'\psi_0},
\end{multline}
averaging over the initial orientation $\phi_0$, and focusing entirely on the dependency on the initial $\psi_0$. The derivatives in $k_x$ and $k_y$ restrict the right-hand side to diagrams with exactly one self-propulsion vertex, \Eref{pertvert1}, sending the $\phi-$mode on the right, $n'=0$, to the $\phi-$mode on the left, $n=\pm1$ respectively. The propagator to the right of the self propulsion vertex contains any number of tumble vertices, causing a 'decrease' in the $\psi-$mode by 0, \Eref{pertvert2}. The propagator to the left of the self-propulsion vertex contains exactly $|m'|$ tumble vertices. For $m'>0$, we need $n'=1$ to bring the $\psi-$mode back down to 0. Similarly for $m'<0$, we need $n'=-1$ to bring to $\psi-$mode up to 0. The contribution from $m'=0$ corresponds to the result integrated over initial $\psi_0$
\begin{multline}\label{pos_orr_cont}
   \langle \chi_{1,0} (\mathbf{k},\omega) \tilde{\chi}_{0,m'}(\mathbf{k}',\omega')\rangle \pm \langle \chi_{-1,0} (\mathbf{k},\omega) \tilde{\chi}_{0,m'}(\mathbf{k}',\omega')\rangle \\ \hat{=}
    \begin{cases}\hspace{0.8em}\begin{tikzpicture}[baseline={([yshift=0.6ex]current bounding box.center)}]
    \begin{feynman}
    \vertex (a) at (-0.15,0);
    \vertex (a1) at (0.4,0);
    \node[circle,
    fill=red,color=red,scale=0.5] (b) at (0.8,0);
    \vertex (b1) at (0.8,-0.15);
    \vertex (c) at (1.2,0);
    \vertex (d) at (1.6,0);
    \node[circle,
    fill=red,color=red,scale=0.5] (e) at (2.0,0);
    \vertex (e2) at (2.0, -0.15);
    \node[isosceles triangle, isosceles triangle apex angle=60,
    draw,
    rotate=90,
    fill=red,color=red,scale=0.5] (e1) at (2.4,0);
    \vertex (f) at (3.2,0);
    \vertex (A) at (0.5,-0.1);
    \vertex (B) at (2.3,-0.1);
    \diagram*{
    (a1) -- [red, line width=0.25mm, -] (c)-- [red, line width=0.25mm, dotted] (d)-- [red, line width=0.25mm, -] (e1)-- [red, line width=0.25mm, gluon](f)};
    \vertex [above=0.2em of a1] {\(1,0\)};
    \vertex [below=0.2em of f] {\(0,m'\)};
    \draw [thick, black,decorate,decoration={brace,amplitude=6pt,mirror}](b1) -- (e2) node[black,midway,yshift=-0.4cm]{\footnotesize $m'$};
    \end{feynman}
    \end{tikzpicture}\hat{=}\hspace{0.5em}\frac{v(k_x-\imag k_y)}{2}\eval{G(\mathbf{k},0,m',\omega)}_{\gamma=0} \gamma^{m'} \prod_{j=0}^{m'} G(\mathbf{k},1,j,\omega) & m'>0 \\ \pm\begin{tikzpicture}[baseline={([yshift=0.6ex]current bounding box.center)}]
    \begin{feynman}
    \vertex (a) at (0.55,0);
    \vertex (a1) at (0.4,0);
    \node[circle,
    fill=red,color=red,scale=0.5] (b) at (0.8,0);
    \vertex (b1) at (0.8,-0.15);
    \vertex (c) at (1.2,0);
    \vertex (d) at (1.6,0);
    \node[circle,
    fill=red,color=red,scale=0.5] (e) at (2.0,0);
    \node[isosceles triangle, isosceles triangle apex angle=60,
    draw,
    rotate=270,
    fill=red,color=red,scale=0.5] (e1) at (2.4,0);
    \vertex (e2) at (2.0, -0.15);
    \vertex (f) at (3.2,0);
    \vertex (A) at (0.5,-0.1);
    \vertex (B) at (2.3,-0.1);
    \diagram*{
    (a1) -- [red, line width=0.25mm, -] (c)-- [red, line width=0.25mm, dotted] (d)-- [red, line width=0.25mm, -] (e1)--[red, line width=0.25mm, gluon](f)};
    \vertex [above=0.2em of a] {\(-1,0\)};
    \vertex [below=0.2em of f] {\(0,m'\)};
    \draw [thick, black,decorate,decoration={brace,amplitude=6pt,mirror}](b1) -- (e2) node[black,midway,yshift=-0.4cm]{\footnotesize $-m'$};
    \end{feynman}
    \end{tikzpicture}\hat{=}\hspace{0.5em} \pm \frac{v(k_x+\imag k_y)}{2}\eval{G(\mathbf{k},0,m',\omega)}_{\gamma=0}\gamma^{-m'} \prod_{j=0}^{-m'} G(\mathbf{k},1,j,\omega)& m'<0\\
    \begin{tikzpicture}[baseline={([yshift=-0.5ex]current bounding box.center)}]
    \begin{feynman}
    \vertex (a) at (0,0);
    \node[isosceles triangle, isosceles triangle apex angle=60,
    draw,
    rotate=90,
    fill=red,color=red,scale=0.5] (b) at (0.8,0);
    \vertex (c) at (1.6,0);
    \diagram*{
    (a)  -- [red, line width=0.25mm, -](b) -- [red, line width=0.25mm, gluon] (c)};
    \vertex [above=0.15em of a] {\(1,0\)};
    \vertex [below=0.15em of c] {\(0,0\)};
    \end{feynman}
    \end{tikzpicture} \pm \begin{tikzpicture}[baseline={([yshift=-0.5ex]current bounding box.center)}]
    \begin{feynman}
    \vertex (a) at (0,0);
    \node[isosceles triangle, isosceles triangle apex angle=60,
    draw,
    rotate=270,
    fill=red,color=red,scale=0.5] (b) at (0.8,0);
    \vertex (c) at (1.6,0);
    \diagram*{
    (a)  -- [red, line width=0.25mm, -](b)-- [red, line width=0.25mm, gluon] (c)};
    \vertex [above=0.15em of a] {\(-1,0\)};
    \vertex [below=0.15em of c] {\(0,0\)};
    \end{feynman}
    \end{tikzpicture} \hat{=}\,\bigl(\frac{v(k_x-\imag k_y)}{2}\pm\frac{v(k_x+\imag k_y)}{2}\bigr)G(\mathbf{k},1,0,\omega)\eval{G(\mathbf{k},0,0,\omega)}_{\gamma=0} & m'=0\end{cases}
\end{multline}
When using \Eref{pos_orr_cont} in \Erefs{xcos} and \eref{ysin}, the derivatives with respect to $k_x$ and $k_y$ will operate on the self propulsion vertex, \Eref{pertvert1}. Their respective effect will render the expression for $\langle x(t) \cos(\phi(t)) \rangle$ identical to that of $\langle y(t) \sin(\phi(t)) \rangle$. After simplifying \Eref{pos_orr_cont} and using the definiton of the bare propagator, \Eref{bareprop}, we can write the cross-correlator as
\begin{multline}
    \langle \mathbf{x}(t)\cdot\mathbf{e}(\phi(t))\rangle = \frac{v}{2}\int\dbar\omega e^{-\imag \omega t} \biggl[\frac{2}{(-\imag\omega + D_\phi +\gamma+r)(-\imag\omega + r)}\biggr. \\ \biggl.+ \sum_{m=1}^{\infty} \frac{\cos{(m\psi_0)\gamma^m}}{-\imag\omega+D_\psi m^2+r}\prod_{i=0}^{m}\frac{1}{-\imag\omega+ D_\phi + D_\psi i^2 + \gamma + r}\biggr].
\end{multline}
\begin{figure}[t!]
\centering
\includegraphics[width=0.6\textwidth]{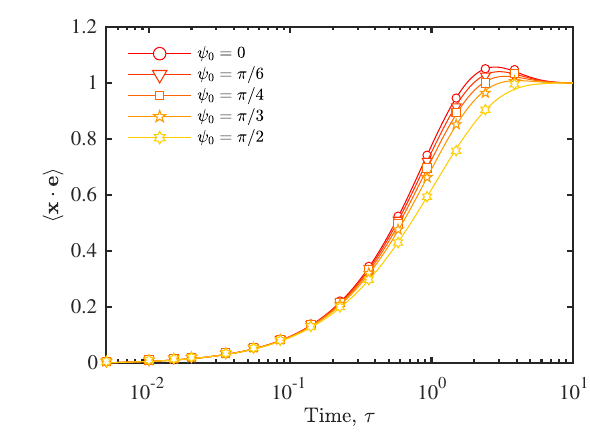}
\caption{Position-orientation cross-correlator --- $\langle\mathbf{x}(t)\cdot\mathbf{e}(\phi(t))\rangle$ plotted as a function of time, $t$, for a range of initial reorientation angles $\psi_0$ with $D_\phi=0, D_\psi=1,\gamma=1$. Symbols indicate numerical simulations, solid lines \Eref{pos_orr2}.}
\label{fig:cross}
\end{figure}
Integrating and taking $r \to 0$ we find
\begin{multline}\label{pos_orr1}
     \langle \mathbf{x}(t)\cdot\mathbf{e}(\phi(t))\rangle = \frac{v}{D_\phi + \gamma}[1-e^{-(D_\phi + \gamma)t}] \\+ v\sum_{m=1}^{\infty} \cos{(m\psi_0)}\gamma^m\biggl[e^{-D_\psi m^2 t}\prod_{i=0}^{m}\frac{1}{D_\phi+D_\psi(i^2 - m^2)+\gamma}\biggr.\\ \biggl. - \sum_{\ell=0}^{m}\frac{e^{-(D_\phi + D_\psi\ell^2 + \gamma)t}}{D_\phi + D_\psi(\ell^2-m^2)+\gamma}\prod_{i=0,i\neq\ell}^{m}\frac{1}{D_\psi(i^2-\ell^2)}\biggr]
\end{multline}
As for \Eref{msdfv2}, we consider $t\to0$, which should produce $\langle \mathbf{x}(t\to0)\cdot\mathbf{e}(\phi(t\to0))\rangle$ = 0. This becomes obvious only after writing \Eref{pos_orr1} as
\begin{multline}\label{pos_orr2}
     \langle \mathbf{x}(t)\cdot\mathbf{e}(\phi(t))\rangle = \frac{v}{D_\phi + \gamma}[1-e^{-(D_\phi + \gamma)t}] \\+ v\sum_{m=1}^{\infty} \cos{(m\psi_0)}\gamma^m\biggl[e^{-D_\psi m^2 t}\sum_{\ell=0}^{m}\frac{1-e^{-(D_\phi+D_\psi(\ell^2-m^2)+\gamma)t}}{D_\phi+D_\psi(\ell^2-m^2)+\gamma}\prod_{i=0,i\neq\ell}^{m}\frac{1}{D_\psi(i^2-\ell^2)}\biggr],
\end{multline}
which also recovers the cross-correlator for an active Brownian particle with additional tumbling by a uniform reorientation angle in the limit $D_\psi \to \infty$, making the second line in \Eref{pos_orr2} vanish. At late-times, the cross-correlator plateaus at $\lim_{t\to\infty}\langle \mathbf{x}(t)\cdot\mathbf{e}(\phi(t))\rangle = v/(D_\phi + \gamma)$, identical to an active Brownian particle with tumbling \cite{Bechinger2016} as $\psi$ fully decorrelates, \Fref{fig:cross}. 
The cross-correlator, \Erefs{pos_orr1} and \eref{pos_orr2}, is symmetric under a change of sign of initial $\psi_0$, i.e. a change in initial tumble direction. To capture the parity of chirality, we calculate the position-perpendicular orientation cross-correlator
\begin{equation}\label{pos_perp}
    \langle \mathbf{x}(t)\cdot\mathbf{e}^{\perp}(\phi(t))\rangle = \langle y(t)\cos{(\phi(t))\rangle - \langle x(t)\sin{(\phi(t))\rangle}},
\end{equation}
an observable that is antisymmetric in initial $\psi_0$, as we demonstrate in the following. The two terms on the right of \Eref{pos_perp} can be expressed as 
\begin{multline}\label{ycos}
    \langle y(t)\cos{(\phi(t)}\rangle = \frac{\imag}{2}\int\dbar\omega e^{-\imag \omega t}\sum_{m'=-\infty}^{\infty} \eval{\partial_{k_y}}_{\mathbf{k}=\mathbf{0}}[\langle \chi_{-1,0} (\mathbf{k},\omega) \tilde{\chi}_{0,m'}(\mathbf{k}',\omega')\rangle\\ + \langle \chi_{1,0} (\mathbf{k},\omega) \tilde{\chi}_{0,m'}(\mathbf{k}',\omega')\rangle]e^{-\imag m'\psi_0},
\end{multline}
and
\begin{multline}\label{xsin}
    \langle x(t)\sin{(\phi(t))}\rangle = \frac{1}{2}\int\dbar\omega e^{-\imag \omega t}\sum_{m'=-\infty}^{\infty} \eval{\partial_{k_x}}_{\mathbf{k}=\mathbf{0}}[\langle \chi_{-1,0} (\mathbf{k},\omega) \tilde{\chi}_{0,m'}(\mathbf{k}',\omega')\rangle\\- \langle \chi_{1,0} (\mathbf{k},\omega) \tilde{\chi}_{0,m'}(\mathbf{k}',\omega')\rangle]e^{-\imag m'\psi_0}.
\end{multline}

To evaluate and simplify \Erefs{ycos} and \eref{xsin} we work along the same lines as above. Notably, the contribution from the incoming $\psi-$mode $m'=0$ vanishes as it corresponds to uniformly distributed initial reorientation angle and thus a lack of chirality. We finally arrive at
\begin{flalign}\label{crossperp}
    \langle \mathbf{x}(t)\cdot\mathbf{e}^{\perp}(\phi(t))\rangle = -v\sum_{m=1}^{\infty} \sin{(m\psi_0)}\gamma^m\biggl[e^{-D_\psi m^2 t}\sum_{\ell=0}^{m}\frac{1-e^{-(D_\phi+D_\psi(\ell^2-m^2)+\gamma)t}}{D_\phi+D_\psi(\ell^2-m^2)+\gamma}\prod_{i=0,i\neq\ell}^{m}\frac{1}{D_\psi(i^2-\ell^2)}\biggr].
\end{flalign}

\begin{figure}[t!]
\centering
\includegraphics[width=0.6\textwidth]{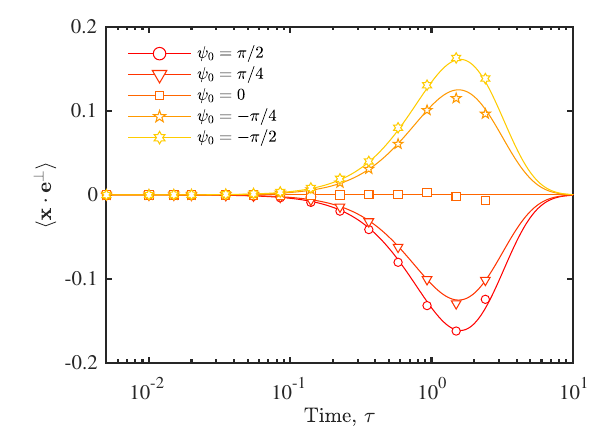}
\caption{Position-perpendicular orientation cross-correlator --- $\langle\mathbf{x(t)}\cdot\mathbf{e}^{\perp}(\phi(t))\rangle$ plotted as a function of time, $t$, for a variety of initial reorientation angles $\psi_0$ with $D_\phi=0, D_\psi=1,\gamma=1$. Symbols indicate numerical simulations, solid lines \Eref{crossperp}.}
\label{fig:crosscorr}
\end{figure}

In \Fref{fig:crosscorr}, we see that stronger correlations arise for $1/(D_{\phi}+\gamma)=\tau_p\ll t \ll \tau_\psi=1/D_\psi$, where tumbles occur before $\psi$ fully decorrelates, resulting in correlated motion. In direct contrast to the position-orientation cross-correlator, \Erefs{pos_orr1} and \eref{pos_orr2}, \Eref{crossperp} displays antisymmetry in $\psi_0$, simply as $\sin(m\psi_0)$ is odd. To leading order, $m=1$, the correlator, \Eref{crossperp}, is negative for $\psi_0 \in [0,\pi)$ and positive otherwise, while it vanishes at large $t$ due to the exponential $\exp{-D_\psi m^2 t}$, thus reflecting the transient nature of the chirality.

\section{Conclusion}\label{sec:conclusion}

In the present work, we have derived a Doi-Peliti field theory to describe the dynamics of transient chiral active particles (TCAPs) in two-dimensions. This allowed us to calculate a range of observables that have in turn allowed us to study memory effects in the reorientation dynamics under this novel model. In particular, we were able to study the mean squared displacement, both averaged over initial reorientation angle $\psi_0$, \Eref{msd_uniform}, and for a fixed $\psi_0$, \Eref{msdf}. We then calculated the reorientation angle probability density function, \Eref{marginal}, which allowed us to quantify the probability of tumbling to the right (equivalently to the left) given initial reorientation angle $\psi_0$, \Eref{eq:transientchiralityevidence}. We finally turned to orientational moments, and were able to calculate the orientational autocorrelator, \Eref{orr_auto}, position-orientation cross-correlator, \Eref{pos_orr2}, and position-perpendicular orientation cross correlator, \Eref{crossperp}.

We found that averaging over initial $\psi_0$ recovers the MSD of an ABP (or RnT) \cite{HowseETAL:2007} with directional diffusivity that is the sum of the angular diffusion $D_\phi$ and the tumble rate $\gamma$. This in turn allowed us to identify short-time translationally diffusive and ballistic behaviour, \Eref{short}, and long-time effective diffusion, \Eref{long}. By fixing initial $\psi_0$ we found a dependence on the diffusion of the reorientation angle $D_\psi$, as well as a difference in the dependence on $D_\phi$ and $\gamma$, where the latter appears as a prefactor $\gamma^m$ which can be interpreted as the number of tumbles that occur. While the short- and long-time behaviour remain unaffected by $D_\psi$, we identify two intermediate parameter regimes with dynamically different behaviour, governed by the speeds at which $\psi$ and $\phi$ decorrelate. If $\psi$ decorrelates faster than $\phi$, the value of $\psi_0$ has no discernible effect on $\phi$ and we recover effectively active Brownian or run-and-tumble behaviour, \Eref{fast}. If $\psi$ decorrelates slower than $\phi$, we see strong dependence on $\psi_0$, with transiently chiral dynamics, \Eref{slow}.

The reorientaion angle probability density function takes on the expected form for diffusion on a ring \cite{Gradshteyn:2007}, \Eref{marginal}, and allowed us to show that initially tumbling to the right (resp. left) increases the likelihood of subsequently tumbling to the right (resp. left), $P(R|\psi_0)$, giving rise to transiently chiral behaviour \Eref{eq:transientchiralityevidence}. We also saw that the period over which a particle remains transiently chiral is governed by $D_\psi/\gamma$; for a right-tumbling particle, $\psi_0\in[0,\pi)$, $P(R|\psi_0)$ decreases to $1/2$ as $D_\psi/\gamma$ increases because $\psi$ decorrelates faster than it is probed (i.e. when a tumble occurs), meaning subsequent tumbles are less likely to have the same chirality, \Fref{fig:transientchirality}. The orientational autocorrelator, \Eref{orr_auto} and \Fref{fig:auto}, displays exponentially decaying behaviour, with dependence on $\psi_0$ based largely on how fast $\psi$ decorrelates. The position-orientation cross-correlator, \Erefs{pos_orr1} and \eref{pos_orr2}, is comprised of the result for an active Brownian or run-and-tumble particle \cite{Bechinger2016} with corrections that are invariant under a change of sign in $\psi_0$, \Fref{fig:cross}. The position-perpedicular orientation cross-correlator, \Eref{crossperp}, is anti-symmetric in a change of sign of $\psi_0$; if $0<\psi_0<\pi$, then $\langle \mathbf{x}\cdot\mathbf{e}^{\perp} \rangle > 0$. Conversely, if $-\pi<\psi_0< 0$, then $\langle \mathbf{x}\cdot\mathbf{e}^{\perp} \rangle < 0$. The cross-correlator allows us to identify the chirality of the particle in the transiently chiral regime, \Fref{fig:crosscorr}.

The present work illustrates how one approaches a field theory in which the particle has additional degrees of freedom, and uses this framework to analyse the motion of such particles. The dynamics of $\psi$ could be further explored and modified, for example allowing $\psi$ to evolve in a harmonic potential which would display transient chirality depending on the diffusivity of $\psi$ and the strength and position of the potential. This work also paves the way to explore the collective behaviour of such particles, with potential to lead to the emergence of global transiently chiral behaviour.

\section*{Acknowledgements}
We thank Kristian Stølevik Olsen for useful discussions. CB was supported by a Roth PhD scholarship funded by the Department of Mathematics at Imperial College London.
\section*{Bibliography}
\bibliography{larticles,lbooks,NewReferences}

\end{document}